\documentclass[aps,prc,showpacs]{revtex4-1}
\usepackage{amsmath} % Enhanced mathematics
\usepackage{amssymb} % Enhanced mathematics
\usepackage{graphics}
\usepackage{epsfig}
\usepackage{slashed}
\newcommand\nn{\nonumber}
\newcommand\ba{\begin{eqnarray}}
\newcommand\ea{\end{eqnarray}}
\newcommand\alb{\begin{align}}
\newcommand\ale{\end{align}}
\newcommand\be{\begin{equation}}
\newcommand\ee{\end{equation}}

\usepackage{longtable}
\usepackage{latexsym}

\usepackage{amssymb}
\usepackage{amsfonts} % for \mathfrak
\usepackage{amsmath}

\begin{document}
\title{Model independent radiative corrections to elastic deuteron-electron scattering}
\author{G.I.~Gakh}

\affiliation{\it NSC ''Kharkov Institute of Physics and Technology'',
Akademicheskaya, 1, 61108 Kharkov,  and \\
V.N.~Karazin Kharkiv National University, 61022 Kharkov, Ukraine}

\author{M.I.~Konchatnij}
\affiliation{\it NSC ''Kharkov Institute of Physics and Technology'',
Akademicheskaya, 1, 61108 Kharkov, and \\
V.N.~Karazin Kharkiv National University, 61022 Kharkov, Ukraine}

\author{N.P.~ Merenkov}
\affiliation{\it NSC ''Kharkov Institute of Physics and Technology'',
Akademicheskaya, 1, 61108 Kharkov,  and \\
 V.N.~Karazin Kharkiv National University, 61022 Kharkov, Ukraine}

\author{Egle~Tomasi-Gustafsson}
\affiliation{\it IRFU, CEA, Universit\'e Paris-Saclay, 91191 Gif-sur-Yvette Cedex, France}

\date{\today}

\hspace{0.7cm}

\begin{abstract}
The differential cross section for  elastic scattering of deuterons
on electrons at rest is calculated taking into account the QED
radiative corrections to the leptonic part of interaction. These
model-independent radiative corrections arise due to emission of the
virtual and real soft and hard photons as well as to vacuum
polarization. We consider an experimental setup where both final
particles are recorded in coincidence and their energies are
determined within some uncertainties. The kinematics, the cross
section, and the radiative corrections are calculated and numerical
results are presented.
\end{abstract}

\maketitle

%%%%%%%%%%%%%%%%%%
\section{Introduction}
%%%%%%%%%%%%%%%%

The polarized and unpolarized scattering of electrons off protons
and light nuclei has been widely studied since these experiments
give information on the internal structure of these particles.

The determination of the proton electromagnetic form factors, at
$Q^2=-q^2\geq 1$ GeV$^2$, from polarization observables showed a
surprising result: the polarized and unpolarized experiments ended
up with inconsistent values of the form factor ratio (see the review
\cite{PBTG15} and references therein). This puzzle has given rise to
many speculations and different interpretations (for example, taking
into account higher order radiative corrections), suggesting further
experiments (see the review \cite{PPV07}).

In the region of small $Q^2$ one can determine the charge radius of
the proton and of the light nuclei ($r_c$), which is one of the fundamental
quantities in physics. The precise knowledge of its value is important
for the understanding of the structure of the nucleon and deuteron
in the theory of strong interactions (QCD) as well as in the
spectroscopy of atomic hydrogen and deuterium.

Recently, the determination of the proton $r_c$ with muonic atoms
lead to the so-called proton radius puzzle. Experiments on muonic
hydrogen by laser spectroscopy measurements lead to the following
result on the proton charge radius: $r_c=0.84087(39)$
fm  \cite{Aet13}. It is one order of magnitude more precise but smaller by seven
standard deviations compared to the average value $r_c=0.8775(51)$
fm which is recommended by the 2010-CODATA review \cite{MNT16}. The
CODATA value is obtained coherently from hydrogen atom spectroscopy
and electron-proton elastic scattering measurements.

While the corrections to the laser spectroscopy experiments seem
well under control in the frame of QED and may be estimated with a
precision better than 0.1\%, in case of electron-proton elastic
scattering the best achieved precision is of the order of few
percent. Different sources of possible systematic errors of the
muonic experiments have been discussed. However, no definite
explanation of this difference has been given yet (see Ref.
\cite{Aet11} and references therein).

The deuteron form factors have been also extensively investigated
during last years. The discussion of the experimental results can be
found in the reviews \cite{S01,GG02,G03}. One can expect that
analogous problems may arise with the
determination of the deuteron charge radius.

The precise knowledge of the deuteron charge radius can give
additional information about the deuteron internal structure. The
authors of Ref. \cite{LDL14} check the contribution from the
different coordinate intervals of the deuteron wave function to the 
radius and found that it was sizable due to the large $r$ region. So,
they concluded that extrapolation of the wave function in the large
distance is of great interest. A new method which allows to fix the
percentage of the elusive D-state probability, $P_D$, from
experiments presented in Ref. \cite{KB16}. It uses the dependence of
the deuteron charge radius, $r_d$, on the deuteron wave function.
Therefore, the precise knowledge of $r_d$ permits to determine $P_D$
more accurately.

The CREMA collaboration has just published a value of the radius
$r_d$ from laser spectroscopy of the muonic deuterium
($\mu$d) \cite{Pet16},
$$ r_d(\mu d) = 2.1256 (8) fm, $$
again more than 7$\sigma$ smaller than the CODATA-2010 value of
$r_d$ \cite{MTN12}
$$ r_d(CODATA-2010) = 2.1424 (21) fm. $$

As was noted in Ref. \cite{PNU16}, the comparison of the new
$r_d(\mu d)$ value with the CODATA-2010 value may be considered
inadequate or redundant, because the CODATA values of $r_d$ and
$r_p$ are highly correlated. A pedagogical description of the method
to extract the charge radius and the Rydberg constant from laser
spectroscopy in regular hydrogen and deuterium atoms is given in
Ref. \cite{PNU16}. The principle of determining the deuteron radius
from deuterium spectroscopy is exactly analogous to the one
described for hydrogen above. However, not all measurements were
done for deuterium \cite{PNU16}.

We propose to use deuteron elastic scattering on atomic
electrons (the inverse kinematics) for a precise measurement of the
deuteron charge radius. The inverse kinematics allows to reach a
very small values of the four-momentum transfer squared.

The inverse kinematics was considered in a number of papers. It was
shown \cite{G96} that the measurement of the spin correlation
parameters (polarized beam and target) in the proton elastic scattering
on atomic electrons can be used for the measurement of high-energy
proton beam polarization. The cross section and polarization
observables for the proton-electron elastic scattering were derived
in a relativistic approach assuming the one-photon-exchange
approximation \cite{GDMTB}. The suggestion to use this reaction for
the determination of the proton charge radius was considered in
\cite{GDTMB}. The model-independent radiative corrections to the
differential cross section for elastic proton-electron scattering
have been calculated in \cite{GKMT} in the case of experimental
setup when both the final particles are recorded in coincidence. The
inverse kinematics was proposed to measure neutron capture cross
section of unstable isotopes \cite{RL13}. For proton and
$\alpha$-induced reactions it was suggested to employ a radioactive
ion beam hitting a proton or helium target at rest.

In this paper, we study the process of the elastic scattering of
deuterons on electrons at rest taking into account the QED radiative
corrections to the leptonic part of interaction. These
model-independent radiative corrections arise due to the emission of
the virtual and real soft and hard photons as well as to the vacuum
polarization. We analyze an experimental setup when the scattered
deuteron and electron are recorded in coincidence and their energies
are determined within some uncertainties. The kinematics, the cross
section, and the radiative corrections are calculated and numerical
results are presented.

\begin{figure}[t]
\centering
\includegraphics[width=0.44\textwidth]{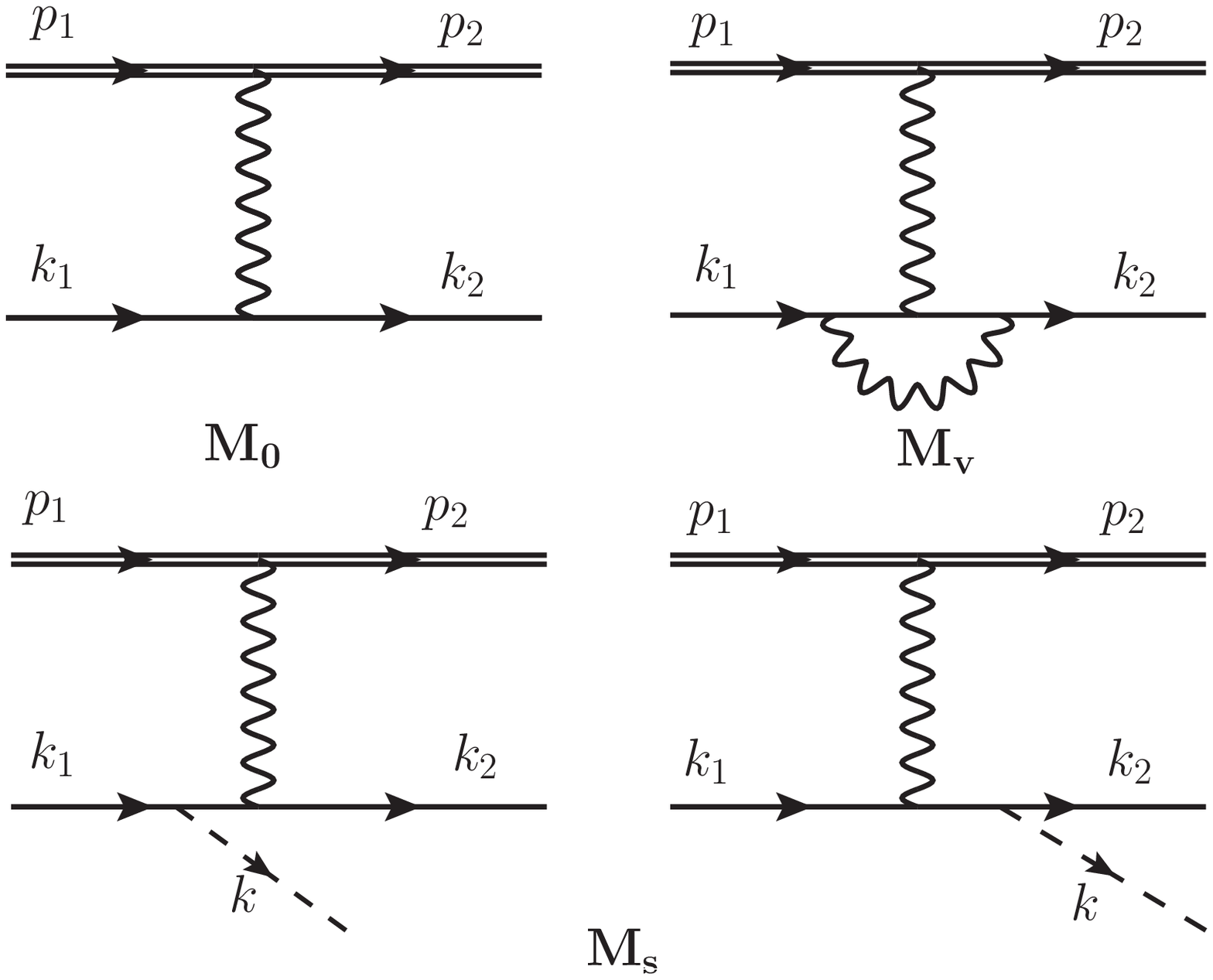}
\includegraphics[width=0.46\textwidth]{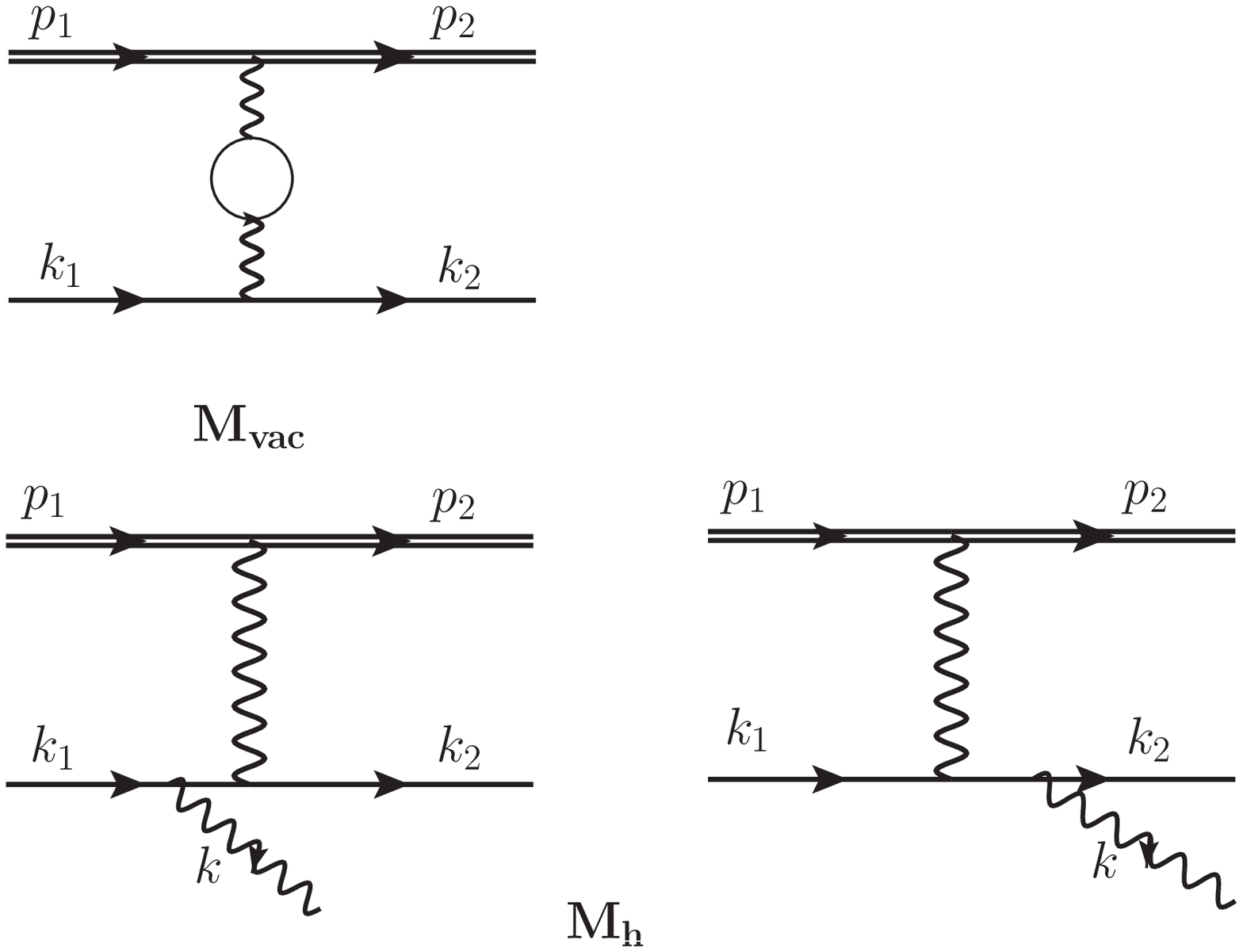}
\caption{Feynman's diagrams corresponding to the Born
 approximation and first order virtual  radiative corrections (top), and to initial and final real photon emission,
  $M_s$ (soft) and $M_h$ (hard), from the lepton vertex (bottom). }
 \label{diagrams}
\end{figure}

%%%%%%%%%%%%%%%%
\section{Formalism}
%%%%%%%%%%%%%%%%
Let us consider the reaction
\begin{equation}\label{eq:1}
d(p_1)+e^-(k_1)\to d(p_2)+e^-(k_2),
\end{equation}
where the particle momenta are indicated in parenthesis, and
$q=k_1-k_2=p_2-p_1$ is the four momentum of the virtual photon.

%%%%%%%%%%%%%%%%%%%%%%%%%%%%%%%%%
\subsection{Inverse kinematics}
%%%%%%%%%%%%%%%%%%%%%%%%%%%%%%%%%

A general characteristic of all reactions of elastic and inelastic
hadron scattering by atomic electrons (which can be considered at
rest) is the small value of the  momentum transfer squared, even for
relatively large energies of the colliding particles. Let us give
details of the order of magnitude and the dependence of the
kinematic variables, as they are very specific for these reactions.
In particular, the electron mass can not be neglected in the
kinematics and dynamics of the reaction, even when the beam energy
is of the order of  GeV.

One can show that, for a given energy of the deuteron beam, the
maximum value of the four momentum transfer squared, in the
scattering on electrons at rest, is
\begin{equation}\label{eq:2}
(Q^2)_{max}=\frac{4m^2|\vec p|^2}{M^2+2mE+m^2},
\end{equation}
where m(M) is the electron (deuteron) mass, $E (\vec{p})$ is the
energy (momentum) of the deuteron beam. Being proportional to the
electron mass squared, the four momentum transfer squared is
restricted to very small values, where the deuteron can be
considered structureless.

The four momentum transfer squared is expressed as a function of the
energy of the scattered electron, $\epsilon_2$, as:
$q^2=(k_1-k_2)^2=2m(m-\epsilon_2)$, where
\begin{equation}\label{eq:3}
\epsilon_2=m\frac{(E+m)^2+|\vec p|^2\cos^2\theta_e}{(E+m)^2-|\vec
p|^2\cos^2\theta_e},
\end{equation}
where $\theta_e$ is the angle between the deuteron beam and the
scattered electron momenta.

From energy and momentum conservation, one finds the following
relation between the angle and the energy of the scattered electron:
\begin{equation}\label{eq:4}
\cos\theta_e=\displaystyle\frac{(E+m)(\epsilon_2-m)} {|\vec p||\vec
{k_2}|},
\end{equation}
where $\vec{k}_2$ is the momentum of the recoil electron and this
formula shows that $\cos\theta_e\geq 0$ (the electron can never be
scattered backward). One can see from Eq.\,(\ref{eq:3}) that, in the inverse
kinematics, the available kinematical region is reduced to small
values of $\epsilon_2$:
\begin{equation}\label{eq:5}
\epsilon_{2max}=m\frac{2E(E+m)+m^2-M^2}{M^2+2mE+m^2},
\end{equation}
which is proportional to the electron mass. From the momentum
conservation, on can find the following relation between the energy
and the angle of the scattered deuteron $E_2$ and $\theta_d$:
\begin{equation}\label{eq:6}
E_2^{\pm}=\frac {(E+m)(M^2+m E)\pm M|\vec p|^2
\cos\theta_d\sqrt{\frac{m^2}{M^2} -\sin^2\theta_d}}{(E+m)^2-|\vec
p|^2 \cos^2\theta_d},
\end{equation}
and this relation shows that, for one deuteron angle, there may be
two values of the deuteron energies, (and two corresponding values
for the recoil-electron energy and angle as well as for the
transferred momentum $q^2$). This is a typical situation when the
center-of-mass velocity is larger than the velocity of the
projectile in the center of mass, where all the angles are allowed
for the recoil electron. The two solutions coincide when the angle
between the initial and final hadron takes its maximum value, which
is determined by the ratio of the electron and scattered hadron
masses $M_h$, $\sin\theta_{h,max}=m/M_h$. One concludes that hadrons
are scattered on atomic electrons at very small angles, and that the
larger is the hadron mass, the smaller is the available angular
range for the scattered hadron.

%%%%%%%%%%%%%%%%%%%%%%%%%%%%%%%%%
\subsection{Differential cross section}
%%%%%%%%%%%%%%%%%%%%%%%%%%%%%%%%%

In the one-photon exchange (Born) approximation, the matrix element
${\cal M}^{(B)}$ of the  reaction (\ref{eq:1}) can be written as:
\begin{equation}\label{eq:7}
{\cal M}^{(B)}=\frac{e^2}{q^2}j_{\mu}J_{\mu},
\end{equation}
where $j_{\mu}(J_{\mu})$ is the leptonic (hadronic) electromagnetic
current. The leptonic current is
\begin{equation}\label{eq:8}
j_{\mu}=\bar u(k_2)\gamma_{\mu} u(k_1),
\end{equation}
where $u(k_{1,2})$ is the spinor of the incoming (outgoing)
electron. Following the requirements of Lorentz invariance, current
conservation, parity and time-reversal invariance of the hadronic
electromagnetic interaction, the general form of the electromagnetic
current for the deuteron (which is a spin-one particle) is fully
described by three form factors. The hadronic electromagnetic
current can be written as:
\begin{equation}\label{eq:9}
J_{\mu}=(p_1+p_2)_{\mu}\left [-G_1(q^2)U_1\cdot
U_2^*+\frac{1}{M^2}G_3(q^2) \left (U_1\cdot q U_2^*\cdot
q-\frac{q^2}{2}U_1\cdot U_2^*\right )\right ]+ \end{equation}
$$+G_2(q^2)\left (U_{1\mu}U_2^*\cdot q -U_{2\mu}^*U_1\cdot q\right
), $$ where $U_{1\mu}$ and $U_{2\mu}$ are the polarization four
vectors for the initial and final deuteron states. The functions
$G_i(q^2)$,\, i=1,\,2,\,3, are the deuteron electromagnetic form factors,
depending only on the virtual photon four momentum squared. Due to
the current hermiticity, these form factors are the real functions
in the region of the space-like momentum transfer.

These form factors are related to the standard deuteron form
factors: $G_C$ (charge monopole) $G_M$ (magnetic dipole) and  $G_Q$
(charge quadrupole) by the following relations:
\begin{equation}\label{eq:10}
G_M(q^2)=-G_2(q^2), \ \ G_Q(q^2)= G_1(q^2)+G_2(q^2)+2G_3(q^2),
\end{equation}
$$G_C(q^2)=\frac{2}{3}\tau \left [G_2(q^2)-G_3(q^2)\right ]+\left
(1+ \frac{2}{3}\tau\right ) G_1(q^2),~\tau=-\frac{q^2}{4M^2}. $$ The
standard form factors have the following normalization:
\begin{equation}\label{eq:11}
G_C(0)=1, \ \ G_M(0)=\frac{M}{m_N}\mu_d, \ \ G_Q(0)=M^2{\cal Q}_d,
\end{equation}
where $m_N$ is the nucleon mass, $\mu_d=0.857 ({\cal Q}_d=0.2857$
fm$^2$) is the deuteron magnetic (quadrupole) moment.

The matrix element squared is written as:
\begin{equation}\label{eq:12}
|{\cal
M}^{(B)}|^2=16\pi^2\frac{\alpha^2}{q^4}L_{\mu\nu}W_{\mu\nu},
\end{equation}
where $\alpha=e^2/(4\pi)=1/137$ is the electromagnetic fine
structure constant. The leptonic $L_{\mu\nu}$ and hadronic
$H_{\mu\nu}$ tensors are defined as
\begin{equation}\label{eq:13}
L_{\mu\nu}=j_{\mu}j_{\nu}^*,~H_{\mu\nu}=J_{\mu}J_{\nu}^*.
\end{equation}
The leptonic tensor $L_{\mu\nu}$ for unpolarized initial and final
electrons (averaging over the initial electron spin) has the form:
\begin{equation}\label{eq:14}
L_{\mu\nu}=q^2g_{\mu\nu}+2(k_{1\mu}k_{2\nu}+k_{1\nu}k_{2\mu}).
\end{equation}

The hadronic tensor $W_{\mu\nu}$ for unpolarized initial and final
deuterons can be written in the standard form, in terms of two
unpolarized structure functions:
\begin{equation}\label{eq:15}
W_{\mu\nu}=\left ( -g_{\mu\nu}+\frac{q_{\mu}q_{\nu}}{q^2}\right )
W_1(q^2)+P_{\mu}P_{\nu} W_2(q^2),
\end{equation}
where $P_{\mu}=(p_1+p_2)_{\mu}/(2\,M)$. Averaging over the spin of the
initial deuteron, the structure functions $W_i(q^2)$, $i=1,2,$ can
be expressed in terms of the electromagnetic form factors as:
\begin{equation}\label{eq:16}
W_1(q^2)=-\frac{2}{3}q^2(1+\tau)G_M^2(q^2),
\end{equation}
$$W_2(q^2)=4\,M^2\left [
G_C^2(q^2)+\frac{2}{3}\tau G_M^2(q^2)+\frac{8}{9} \tau^2
G_Q^2(q^2)\right ]. $$

The expression of the differential cross section, as a function of
the recoil-electron energy $\epsilon_2$, for unpolarized
deuteron-electron scattering can be written as:
\begin{equation}\label{eq:17}
\frac{d\sigma^{(B)}}{d\epsilon_2}=\frac{\pi\alpha^2}{m|\vec
p|^2}\frac{\cal D}{q^4},
\end{equation}
with
\begin{equation}\label{eq:18}
{\cal D}=2\bigl [M^2q^2+2mE(2mE+q^2)\bigr ]\bigl
[G_C^2(q^2)+\frac{8}{9}\tau^2G_Q^2(q^2)\bigr ]+
\end{equation}
$$+\frac{4}{3}\tau\bigl [4m^2(E^2-M^2)+q^2(m^2-M^2-2\tau
M^2+2mE)\bigr ]G_M^2(q^2). $$ This expression is valid in the
one-photon exchange (Born) approximation in the reference system
where the target electron is at rest.

The expression of the differential cross section, as a function of
the four-momentum transfer squared, is
\begin{equation}\label{eq:19}
\frac{d\sigma^{(B)}}{dq^2}=\frac{\pi\alpha^2}{2m^2|\vec
p|^2}\frac{\cal D}{q^4}.
\end{equation}

And at last, the differential cross section over the
scattered-electron solid angle has the following expression
\begin{equation}\label{eq:20}
\frac{d\sigma^{(B)}}{d\Omega_e}=\frac{\alpha^2}{8m^4|\vec{p}|} \left
(1-\frac{4m^2}{q^2}\right )^{3/2}\frac{\cal D}{E+m}.
\end{equation}

%%%%%%%%%%%%%%%%%%%%%%%%%%%%%%%%%
\section{Radiative corrections}
%%%%%%%%%%%%%%%%%%%%%%%%%%%%%%%%%
Let us consider the model-independent QED radiative corrections
which are due to the vacuum polarization and emission of the virtual
and real (soft and hard) photons in the electron vertex. The
corresponding diagrams are shown in Fig. \ref{diagrams}.

%%%%%%%%%%%%%%%%%%%%%
\subsection{Soft photon emission}
%%%%%%%%%%%%%%%%%%%%%
In this section we give the expressions for the contribution to the
radiative corrections of the soft photon emission  when the photons
are emitted by the initial and final electrons
\begin{equation}\label{eq:21}
d(p_1)+e(k_1)\to d(p_2)+e(k_2)+\gamma (k).
\end{equation}

The matrix element in this case (the photon emitted from the
electron vertex) is given by
\begin{equation}\label{eq:22}
{\cal
M}^{(\gamma)}=\frac{1}{q^2}(4\pi\alpha)^{3/2}j_{\mu}^{(\gamma)}J_{\mu},
\end{equation}
where the electron current corresponding to the photon emission is
\begin{equation}\label{eq:23}
j_{\mu}^{(\gamma)}=\bar{u}(k_2)\left
[\frac{1}{d_1}\gamma_{\mu}(\hat{k}_1-\hat{k}+m)
\gamma_{\rho}+\frac{1}{d_2}\gamma_{\rho}(\hat{k}_2+\hat{k}+m)
\gamma_{\mu}\right ]u(k_1)A^*_{\rho},
\end{equation}
where $A_{\rho}$ is the polarization four vector of the emitted
photon and $d_1=-2k\cdot k_1, d_2=2k\cdot k_2$.

The differential cross section of reaction (\ref{eq:2}1) can be written as
\begin{equation}\label{eq:24}
d\sigma^{(\gamma)}=\frac{(2\pi)^{-5}}{32m|\vec{p}|}|{\cal
M}^{(\gamma)}|^2 \frac{d^3\vec k_2}{\epsilon_2}\frac{d^3\vec
p_2}{E_2}\frac{d^3\vec k}{\omega} \delta^4(k_1+p_1-k_2-p_2-k).
\end{equation}

It is necessary to integrate over the photon phase space. Since the
photons are assumed to be soft,  then the integration over the
photon energy is restricted to $|\vec{k}|\leq \bar\omega$. The
quantity $\bar\omega$ is determined by particular experimental
conditions and it is assumed that $\bar\omega$ is sufficiently small
to neglect the momentum $k$ in the $\delta $ function and in the
numerators of the matrix element of the process (\ref{eq:21}). In order to
avoid the infrared divergence, which occurs in the soft photon cross
section, a small fictitious photon mass $\lambda $ is introduced.

The differential cross section of the process (\ref{eq:21}), integrated over
the soft photon phase space, can be written as
\begin{equation}\label{eq:25}
d\sigma^{(soft)}=\delta_sd\sigma^{(B)},
\end{equation}
where the radiative correction due to the soft photon emission is
%\cite{GKM}
\begin{eqnarray}
\delta_s&=&\frac{\alpha}{\pi}\bigl
\{1-2\ln{\frac{2\bar\omega}{\lambda}}+\frac{\epsilon_2}{|\vec{k}_2|}\bigl
[\ln{\frac{\epsilon_2+|\vec{k}_2|}{m}} \bigl (
1+2\ln{\frac{2\bar\omega}{\lambda}}
+\ln{\frac{\epsilon_2+|\vec{k}_2|}
{m}}+2\ln{\frac{m}{2|\vec{k}_2|}}\bigr )-
\nn\\
&&-\frac{\pi^2}{6}+Li_2\bigl
(\frac{\epsilon_2-|\vec{k}_2|} {\epsilon_2+|\vec{k}_2|}\bigr )\bigr
]\bigr  \}, \label{eq:26}
\end{eqnarray}
where $Li_2(x)$ is the Spence (dilogarithm) function
defined as
$$ Li_2(x)=-\int_0^x\frac{\ln{(1-t)}}{t}dt. $$

%%%%%%%%%%%%%%%%%%%
\subsection{Virtual photon emission}
%%%%%%%%%%%%%%%%%%%
In this section, we give the expressions for the contribution to the
radiative corrections of the virtual photon emission in the electron
vertex (the electron vertex correction) and the vacuum polarization
term.

As we limit ourselves to the calculation of the radiative
corrections at the order of $\alpha  $ in comparison with the Born
term, it is sufficient to calculate  the interference of the Born
matrix element with ${\cal M}^{(virt)} $
\begin{equation}\label{eq:27}
|{\cal M}|^2=|{\cal M}^{(B)}|^2+2Re[{\cal M}^{(virt)}{\cal
M}^{(B)*}]=(1+\delta_1+\delta_2)|{\cal M}^{(B)}|^2,
\end{equation}
where the term $\delta_1$ is due to the modification of the
$\gamma_{\mu}$ term in the electron vertex, and the term $\delta_2$
is due to the presence of the $\sigma_{\mu\nu}q_{\nu}$ structure in
the electron vertex.

The radiative corrections due to the emission of the virtual photon
in the electron vertex can be written as
\begin{eqnarray}
\delta_1&=& \frac{\alpha}{\pi}\bigl \{-2+2\ln\frac{m}{\lambda}\bigl
[1-\frac{\epsilon_2}{|\vec{k}_2|}\ln\bigl
(\frac{\epsilon_2+|\vec{k}_2|}{m} \bigr )\bigr
]+\frac{m+3\epsilon_2}{2|\vec{k}_2|}\ln
\bigl(\frac{\epsilon_2+|\vec{k}_2|}{m}\bigr )-
\frac{1}{2}\frac{\epsilon_2}{|\vec{k}_2|}\ln\bigl
(\frac{Q^2}{m^2}\bigr )
\ln\bigl(\frac{\epsilon_2+|\vec{k}_2|}{m}\bigr )+
\nn\\
&&+\frac{\epsilon_2}{|\vec{k}_2|}
\bigl [-\ln\bigl (\frac {m+\epsilon_2}{|\vec{k}_2|}\bigr ) \ln\bigl
(\frac{\epsilon_2+|\vec{k}_2|}{m}\bigr )+ Li_2\bigl
(\frac{\epsilon_2+|\vec{k}_2|+m}{2\left (m+\epsilon_2\right )}\bigr
)- Li_2\bigl (\frac{\epsilon_2-|\vec{k}_2|+m}{2\left
(m+\epsilon_2\right )}\bigr ) \bigr ]\bigr \}, \\
\label{eq:28}
\delta_2&=&4\frac{\alpha}{\pi}\frac{mM^2q^2}{|\vec{k}_2|D}(1+\tau
)\ln\bigl (\frac{\epsilon_2 +|\vec{k}_2|}{m}\bigr )
(G_C^2-\frac{4}{3}\tau G_M^2+\frac{8}{9}\tau^2G_Q^2).
\label{eq:29}
\end{eqnarray}

The radiative correction due to the vacuum polarization is (the
electron loop has been taken into account):
\begin{equation}\label{eq:30}
\delta^{(vac)}=\frac{2\alpha}{3\pi}\left\{
-\frac{5}{3}+4\frac{m^2}{Q^2}+
(1-2\frac{m^2}{Q^2})\sqrt{1+4\frac{m^2}{Q^2}}\ln\frac{\sqrt{1+4\frac{m^2}{Q^2}}+1}
{\sqrt{1+4\frac{m^2}{Q^2}}-1}\right\}.
\end{equation}
For small and large values of the $Q^2$ variable we have
$$
\mbox{If \ \ } Q^2\ll m^2,   \ \ \
\delta^{(vac)}=\frac{2\alpha}{15\pi}\frac{Q^2}{m^2}, $$ $$ \mbox{If
\ \ }  \ \  Q^2\gg m^2,   \ \ \
\delta^{(vac)}=\frac{2\alpha}{3\pi}\left
[-\frac{5}{3}+\ln\frac{Q^2}{m^2}\right ]. $$

Taking into account the radiative corrections given by Eqs. (\ref{eq:26}, \ref{eq:28},
\ref{eq:29}, \ref{eq:30}), we obtain the following expression for the differential
cross section:
\begin{equation}\label{eq:31}
d\sigma^{(RC)}=(1+\delta_0+\bar\delta +\delta^{(vac)})d\sigma^{(B)},
\end{equation}
where the radiative corrections $\delta_0$ and $\bar\delta $ are
given by
\begin{eqnarray}\label{eq:32}
\delta_0&=&\frac{2\alpha}{\pi}\ln\frac{\bar\omega}{m}\bigl
[\frac{\epsilon_2}{|\vec{k}_2|}\ln\bigl
(\frac{\epsilon_2+|\vec{k}_2|} {m}\bigr )-1\bigr ],\nn\\
\bar\delta &=&\frac{\alpha}{\pi} \Bigl \{ -1-2\ln2+
\frac{\epsilon_2}{|\vec{k}_2|} \Bigl [\ln \left(
\frac{\epsilon_2+|\vec{k}_2|}{m}\right ) \Bigl (1+ \ln \left (
\frac{\epsilon_2+|\vec{k}_2|}{m} \right ) +2\ln
\left(\frac{m}{|\vec{k}_2|}\right )+ \frac{m+3\epsilon_2}
{2\epsilon_2}-
\nn\\
&&
 -\ln\left
(\frac{\epsilon_2+m}{|\vec{k}_2|}\right)- \frac{1}{2}\ln\left
(\frac{Q^2}{m^2}\right )\Bigr )+4m\frac{M^2q^2}{\epsilon_2{\cal D}} (1+\tau
)\ln\bigl (\frac{\epsilon_2 +|\vec{k}_2|}{m}\bigr )
(G_C^2-\frac{4}{3}\tau G_M^2+\frac{8}{9}\tau^2G_Q^2)- \nn\\
&&
-\frac{\pi^2}{6}+Li_2\left
(\frac{\epsilon_2-|\vec{k}_2|}{\epsilon_2+|\vec{k}_2|}\right)+
Li_2\left(\frac{\epsilon_2+|\vec{k}_2|+m}{2(\epsilon_2+m)}\right)-
Li_2\left(\frac{\epsilon_2-|\vec{k}_2|+m}{2(\epsilon_2+m)}\right
)\Bigr ]\Bigr \}.
\label{eq:33}
\end{eqnarray}
We separate the contribution $\delta_0$ since it can be summed up in
all orders of the perturbation theory using the exponential form of
the electron structure functions \cite{KF85}. To do this it is
sufficient to keep only the exponential contributions in the
electron structure functions. The final result can be obtained by
the substitution of the term $(1+\delta_0 )$ by the following term
\begin{equation}\label{eq:34}
\left(\displaystyle\frac{\bar\omega}{m} \right)^{\beta}
\displaystyle\frac{\beta}{2}\int_0^1x^{\frac{\beta}{2}-1}(1-x)^{\frac{\beta}{2}}dx,
\end{equation}
where $$ \beta =\displaystyle\frac{2\alpha}{\pi} \left
[\frac{\epsilon_2}{|\vec{k}_2|}\ln \left (\displaystyle\frac{\epsilon_2+
|\vec{k}_2|}{m}\right)-1 \right ]. $$

%%%%%%%%%%%%%%%%%%%
\subsection{Hard photon emission}
%%%%%%%%%%%%%%%%%%%

In this section we calculate the radiative correction due to the
hard photon emission (with the photon energy $\omega>\bar{\omega}$) from 
the initial and recoil electrons only (the
model-independent part). The contribution due to radiation from the
initial and scattered deuterons (the model-dependent part) requires
a special consideration and we leave it for other investigations. We
consider the experimental setup when only the energies of the
scattered deuteron and final electron are measured.

The differential cross section of the reaction (\ref{eq:21}), averaged over
the initial particle spins, can be written as
\begin{equation}\label{eq:35}
d\,\sigma^{(h)}=\frac{\alpha^3}{32\,\pi^2}\,\frac{1}{m\,|\vec p|}\,
\frac{L_{\mu\nu}^{(\gamma)}\,W_{\mu\nu}}{q_1^4}\frac{d^3k_2}{\epsilon_2}
\frac{d^3p_2}{E_2}\frac{d^3k}{\omega} \delta(p_1+k_1-p_2-k_2-k),
\end{equation}
where $ q_1=k_1-k_2-k$ and the leptonic tensor has the following
form
\begin{eqnarray}
L_{\mu\nu}^{(\gamma)}&=&A_0\tilde{g}_{\mu\nu}+A_1\tilde{k}_{1\mu}\tilde{k}_{1\nu}+
A_2\tilde{k}_{2\mu}\tilde{k}_{2\nu}+A_{12}(\tilde{k}_{1\mu}\tilde{k}_{2\nu}+
\tilde{k}_{1\nu}\tilde{k}_{2\mu}), \nn\\
A_0&=&4\left [\frac{d_1}{d_2}+\frac{d_2}{d_1}-2q_1^2\left
(\frac{m^2}{d_1^2}+\frac{m^2}{d_2^2}+2\frac{k_1\cdot
k_2}{d_1d_2}\right )\right ],\
A_1=16\frac{q_1^2}{d_1d_2}-32\frac{m^2}{d_2^2}, \nn\\
 A_2&=&16\frac{q_1^2}{d_1d_2}-32\frac{m^2}{d_1^2}, \
A_{12}=-32\frac{m^2}{d_1d_2}\,, \ \ \tilde{g}_{\mu\nu}=g_{\mu\nu}-\frac{q_{1\mu}\,q_{1\nu}}{q_1^2}\,, \ \ \tilde{k}_{i\mu}=k_{i\mu}-\frac{(k_i\,q_1)q_{1\mu}}{q_1^2}\,, \ i=1,\,2. 
\label{eq:36}
\end{eqnarray}

\begin{figure}[t]
\centering
\includegraphics[width=0.4\textwidth]{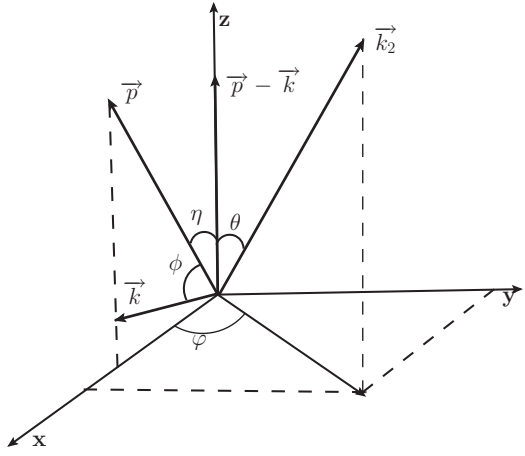}
\caption{Coordinate system and definition of the angles
used for the integration over the variables of the final state.}
\label{Fig:fig2}
\end{figure}

The hadronic tensor is defined by Eqs. (\ref{eq:15}\,,\ref{eq:16}) with the
substitution $q\to q_1.$ The contraction of the leptonic and hadronic tensors reads
\begin{equation}\label{eq:37}
L_{\mu\nu}^{(\gamma)}W_{\mu\nu}=-W_1(q_1^2)\,S_1
+\frac{W_2(q_1^2)}{M^2}\,S_2\,,
\end{equation}
where the functions $S_{1,2}$ have the following expressions
\begin{equation}\label{eq:38}
S_1=8\left (\frac{d_1}{d_2}+\frac{d_2}{d_1}\right
)-\frac{16}{d_1d_2}(2m^2+q_1^2) \left [2k_1\cdot k_2+m^2\left
(\frac{d_1}{d_2}+\frac{d_2}{d_1}\right ) \right]\,,
\end{equation}
\begin{eqnarray}
S_2&=&4M^2\left [\frac{d_1}{d_2}+\frac{d_2}{d_1}-2q_1^2\left
(\frac{m^2}{d_1^2}+\frac{m^2}{d_2^2}+2\frac{k_1\cdot
k_2}{d_1d_2}\right )\right  ]+32\frac{m^2}{d_1d_2}(k\cdot p_1)^2+
16\frac{(k_1\cdot p_1)^2}{d_1}+\nn\\
&&+16\frac{(k_2\cdot p_1)^2}{d_2}+16k_1\cdot p_1k_2\cdot p_1
\left [\frac{1}{d_1}+\frac{1}{d_2}-2\left
(\frac{m^2}{d_1^2}+\frac{m^2}{d_2^2}+2\frac{k_1\cdot
k_2}{d_1d_2}\right  )\right ]+ \nn\\
&&
+16k\cdot p_1\left  [\frac{k_1\cdot p_1}{d_2^2}(d_2-2m^2)-
\frac{k_2\cdot p_1}{d_1^2}(d_1-2m^2)+2\frac{k_1\cdot k_2}{d_1d_2}
(k_2\cdot p_1-k_1\cdot p_1)\right ]. 
\label{eq:37}
\end{eqnarray}
Integrating over the
scattered deuteron variables we obtain the following expression for
the differential cross section
\begin{equation}\label{eq:40}
d\sigma^{(h)}=\frac{\alpha^3}{32\pi^2}\frac{1}{m\,|\vec p|} \int
\frac{d^3{k}}{\omega}\int \frac{d^3{k}_2}{\epsilon_2\,E_2}
\frac{1}{q_1^4}L_{\mu\nu}^{(\gamma)}W_{\mu\nu}\delta(m+E-\epsilon_2-E_2-\omega)\,.
\end{equation}
To integrate further we have to define the coordinate system.
Following Ref. \cite{K64}, where the $\pi-e^-$ scattering has been
analyzed, let us take the $z$-axis along the vector
$\vec{p}-\vec{k}$ and the momenta of the initial deuteron and
emitted photon lie in the $xz$ plane. The momentum of the scattered
electron is defined by the polar $\theta $ and azimuthal $\varphi $
angles as it is shown in Fig. \ref{Fig:fig2}. The angle $\eta (\phi)$ is the
angle between the beam direction and $z$ axis (emitted photon
momentum).

\begin{figure}[t]
\centering
\includegraphics[width=0.4\textwidth]{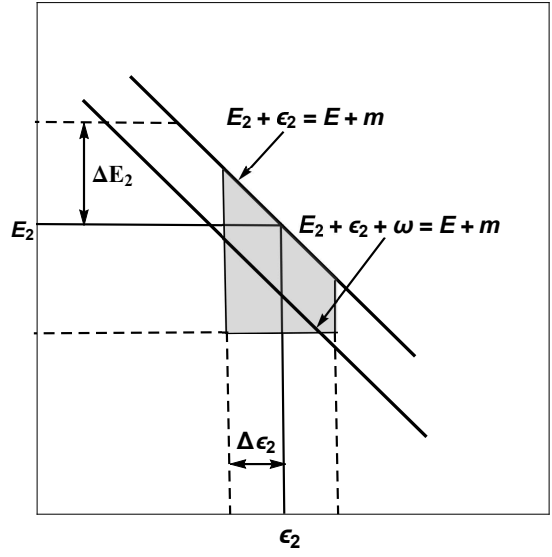}
\caption{Plane of the $E_2$ and $\epsilon_2$
variables where the shaded area represents the kinematically allowed region within the experimental set-up.}
\label{Fig:fig3}
\end{figure}

Integrating over the polar angle of the scattered electron we
obtain:
\begin{equation}\label{eq:41}
\frac{d\sigma^{(h)}}{d\epsilon_2}=\frac{\alpha^3}{32\pi^2}\frac{1}{m\,|\vec
p|}
 \int \frac{d^3\,k}{\omega}\int \frac{d\varphi}{|\vec{p}-\vec{k}|}\,
\frac{1}{q_1^4}L_{\mu\nu}^{(\gamma)}\,W_{\mu\nu}.
\end{equation}
The region of allowed photon momenta should be determined. The
experiment counts those events which, within the accuracy of the
detectors, are considered "elastic". We refer to the experimental
situation where the energies of the scatted deuteron and recoil
electron are measured. Because of the uncertainties in determination
of the recoil electron ($\Delta \epsilon_2$) and scattered deuteron
($\Delta E_2$) energies, which are usually proportional to
$\epsilon_2$ and $E_2,$ respectively, the elastic deuteron-electron
scattering is always accompanied by  hard photon emission with energies up to 
$\Delta \epsilon_2+\Delta E_2$. For deuteron beam
energies of the order of 100 GeV this value can reach a few GeV. The
events corresponding to scattered deuteron energy $E_2\pm \Delta
E_2$ and recoil electron energy  $\epsilon_2 \pm \Delta
\epsilon_2$ (they satisfy the condition $E+m=E_2+\epsilon_2$) are
considered as true elastic events. Here, $\Delta E_2$ and $\Delta
\epsilon_2$ are the uncertainties of the measurement of the final deuteron
and recoil electron energies. The plot of the variable $E_2$ versus
the variable $\epsilon_2$ is shown in Fig. \ref{Fig:fig3}. The shaded area in
this figure represents the region where events are allowed by the experimental
limitations. The relation between the energies $E_2$ and
$\epsilon_2$, as it is shown in Fig. \ref{Fig:fig3}, has to be transformed into a
limit on the possible photon momentum $\vec{k}$.

As we already mentioned, usually the uncertainties $\Delta E_2$ and $\Delta \epsilon_2$ are proportional to $E_2$
and $\epsilon_2,$ respectively. For the deuteron beam energies up to 500\,GeV, the recoil electron energy is 
about two orders of magnitude smaller than the scattered deuteron one. Therefore,  $\Delta E_2\gg\Delta \epsilon_2,$ holds 
and the effect due to nonzero value of $\Delta \epsilon_2$ is negligible. In our further numerical calculations we take $\Delta \epsilon_2=0.$

We consider the
experimental setup where no angles are measured and, therefore, the
orientation of the photon momentum $\vec{k}$ is not limited
and investigate both cases: {\it i})\,$\epsilon_2<\epsilon_{2max}-\Delta E,$
and {\it ii})\,$\epsilon_2>\epsilon_{2max}-\Delta E$, where $\Delta E =\Delta
E_2$ and $\epsilon_{2max}$ is defined by Eq. (\ref{eq:5}).

\begin{figure}[t]
\centering
\includegraphics[width=0.4\textwidth]{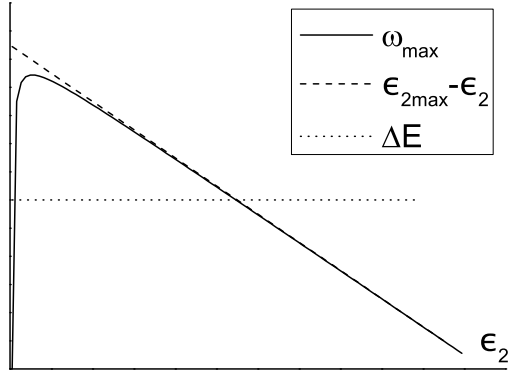}
\caption{The maximum energy of the photon $\omega_{max}$ in the case $\epsilon_2>\epsilon_{2max}-\Delta E$ as given
 by Eq.\,(\ref{eq:42})}.
 \label{Fig:fig4}
\end{figure}

In the first case we get, as experimental limit, the isotropic
condition
$ \omega \leq\Delta E\,,$ whereas in the second case the upper limit of $\omega$ depends on the recoil electron energy
$\omega \leq\omega_{max}\,,$ as it is shown in Fig. \ref{Fig:fig4}.

\begin{figure}[t]
\centering
\includegraphics[width=0.35\textwidth]{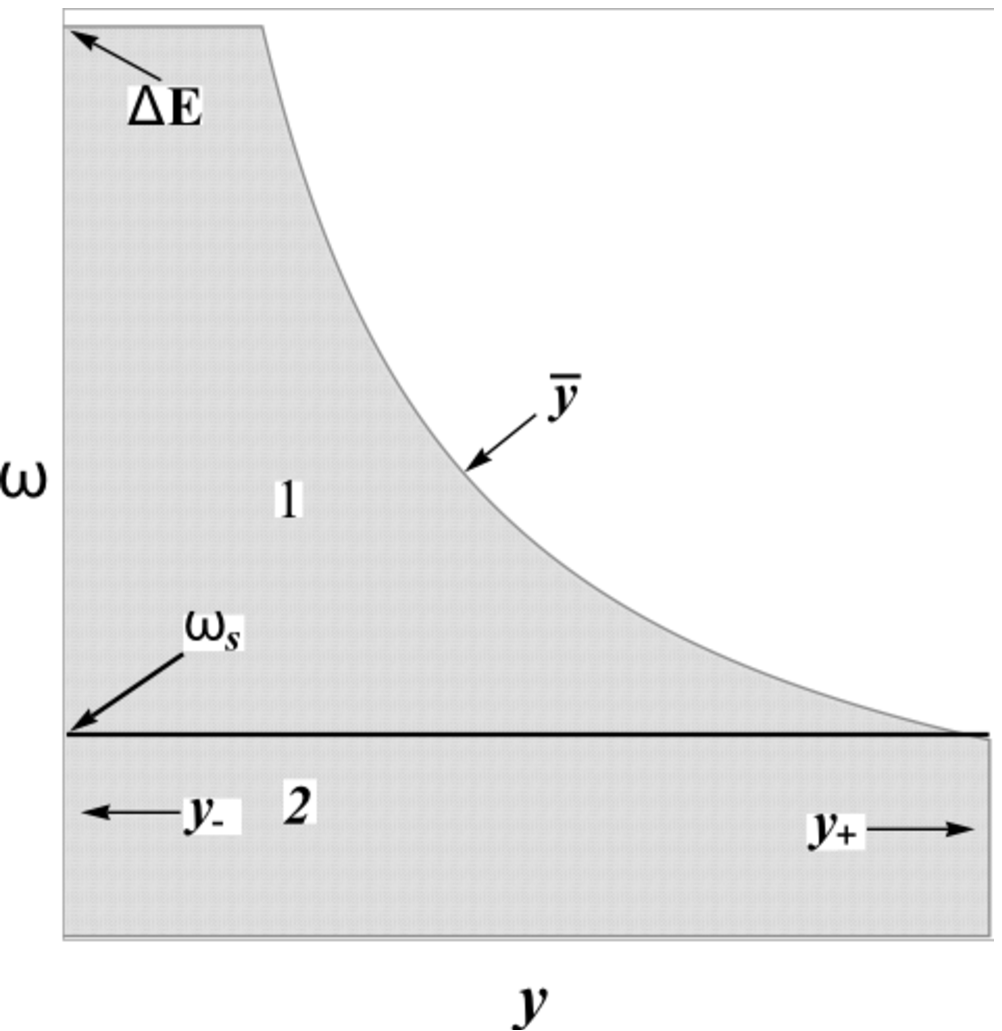}
\hspace{0.5cm}
\includegraphics[width=0.4\textwidth]{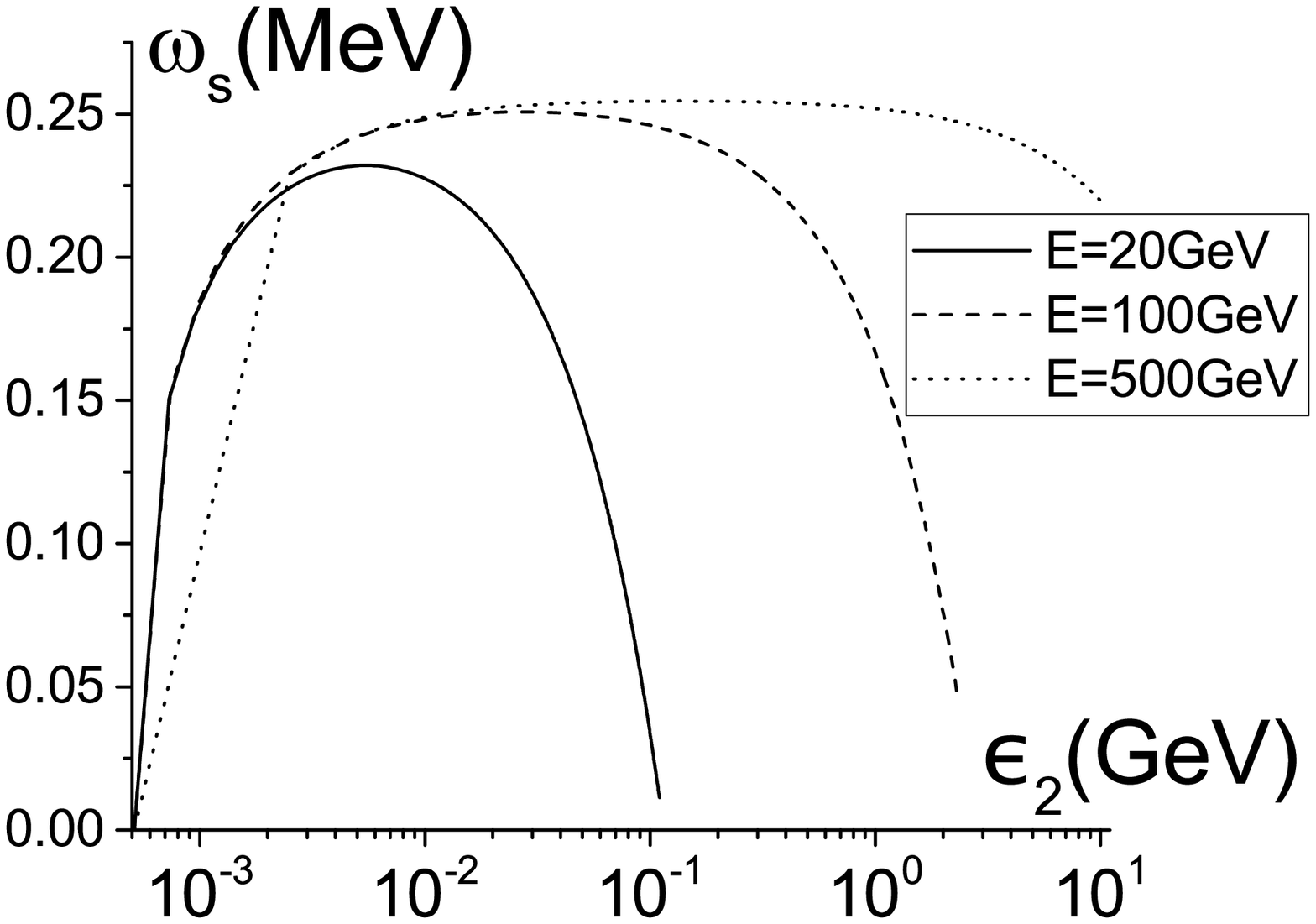}
\caption{In the left panel the integration region over the variables $\omega$
and $y$ in the case $\epsilon_2<\epsilon_{2max}-\Delta E$. Here $y_{\pm}=E\pm p, ~
\bar{y}=[(m-\epsilon_2)(E-\epsilon_2-\omega
)+\sqrt{\epsilon_2^2-m^2}\sqrt{(E+m-\epsilon_2-\omega )^2-M^2}]/
\omega .$ The quantity $\omega_s$ is defined by positive solution of the equation $\bar{y}=y_+$ and given by Eq. (\ref{eq:44}).
 It is shown in the right panel as a function of the recoil electron energy for the different deuteron energies.}
 \label{Fig:fig5}
\end{figure}

Note that quantity $\omega_{max}$ is the root of the equation $y_-=\bar{y}$ and has the following form
\begin{eqnarray}
\omega_{max}&=&\frac{\beta\,A}{B^2-C^2}\,, \ \ \beta=2 m E+m^2+M^2\,,\nn\\
A&=&|\vec{k}_2|\big\{M^2\big[|\vec{p}|(E_0+|\vec{p}|)+\epsilon_-(E-m-2 \epsilon_2)\big]+2 m \epsilon_-(2 E E_0-m \epsilon_-)\big\}-\nn\\
&&-\epsilon_-\big\{(E_0+|\vec{p}|)\big[4 m E E_0-2 m^2 \epsilon_-+M^2(E-\epsilon_2)\big]-M^2 E_0 \epsilon_+\big\}\,,\nn\\
B&=&E(4 m^2+M^2)+m(2 E^2+2 m^2+M^2)-2\beta\,\epsilon_2\,,\ \ \nn\\
C&=&|\vec{p}|(2 m E+M^2)\,, \ \epsilon_{\pm}=\epsilon_2\pm m\,, E_0=E+m-\epsilon_2\,.
\label{eq:42}
\end{eqnarray}
In the chosen coordinate system the element of solid angle becomes:
$d^3 k\to 2\pi\,\omega^2\,d\omega\,d\cos{\phi}$. We introduce a
new variable $y=E-|\vec p|\cos{\phi}>0$ and rewrite Eq.\,(\ref{eq:41}) as
\begin{equation}\label{eq:43}
\frac{d\sigma^{(h)}}{d\epsilon_2}=\frac{\alpha^3}{16\,\pi}\frac{1}{m\,|\vec
p|^2}\int \omega\,d\omega\int d\,y\int\limits_0^{2\pi}
\frac{1}{q_1^4\,|\vec{p}-\vec{k}|}
\Big(-W_1(q_1^2)\,S_1+\frac{W_2(q_1^2)}{M^2}\,S_2\Big)d\,\varphi\,,
\end{equation}
where the integration region over the variables $\omega$ and $y$ for the case $\omega \leq\Delta E$ is
shown in the left panel of Fig. \ref{Fig:fig5}, and
\begin{equation}\label{eq:44}
\omega_s=(|\vec{k}_2|-|\vec{p}|+E_2)\frac{M^2|\vec{k}_2|(|\vec{k}_2|+|\vec{p}|)
-\epsilon_-[M^2(E_2-m)+2m(2EE_2+m^2-m\epsilon_2)]}{M^4-4\epsilon_-[E_2(M^2+m^2+2mE)-
m(M^2+mE)]}\,.
\end{equation}
The quantity $\omega_s$ represents the maximal energy, when the
photon can be emitted in the whole angular phase space. The
dependence of this quantity on the recoil electron energy, at
different values of the deuteron beam energy, is shown in the right panel of Fig.\ref{Fig:fig5}. We
see that it is of the order of the electron mass $m$ in a wide range
of the energies $\epsilon_2$ and $E$.  Because our analytical
calculations for the soft photon correction were performed under the
condition $\bar\omega\ll m,$ where $\bar{\omega}$ is the maximal
energy of the soft photon, we can not identify $\omega_s$ with
$\bar\omega\ll m,$ as it has been done in the paper \cite{K64}.

So, the expression for the cross section given by Eq.\,(\ref{eq:43}) can be
written as a sum of two terms
\begin{equation}\label{eq:45}
\frac{d\sigma^{(h)}}{d\epsilon_2}=\frac{\alpha^3}{16\,\pi}\frac{1}{m\,|\vec
p|^2}\bigg[\int \limits_{\omega_s}^{\Delta
E}\,C_1(\omega)\,d\omega +\int
\limits_{\bar{\omega}}^{\omega_s}C_2(\omega)\,d\omega\bigg]\,,
\end{equation}
where

\begin{eqnarray}
C_1(\omega)&=& \int\limits_{y_-}^{\bar{y}}\int\limits_0^{2\pi}\left [
\frac{\omega}{q_1^4\,|\vec{p}-\vec{k}|}
\left(-W_1(q_1^2)\,S_1+\frac{W_2(q_1^2)}{M^2}\,S_2\right)\right
]d\varphi dy,
\nn\\
C_2(\omega)&=&\int\limits_{y_-}^{y_+}\int\limits_0^{2\pi}\left[
\frac{\omega}{q_1^4\,|\vec{p}-\vec{k}|} \left
(-W_1(q_1^2)\,S_1+\frac{W_2(q_1^2)}{M^2}\,S_2\right )\right]d\varphi dy. 
\label{eq:46}
\end{eqnarray}

The scalar products of various 4-momenta, which enter in the
expressions for $S_1, S_2$ and $q_1^2$, are expressed, in terms of
the angles, as illustrated in Fig. \ref{Fig:fig2}, and the photon energy, as
follows: 
\begin{eqnarray}
d_1&=&-2m\,\omega\,, \ \ k_1\cdot k_2=m\,\epsilon_2\,,
 \ k_1\cdot p_1=m\,E\,, \ \ k\cdot p_1=\omega(E-|\vec
p|\cos{\phi})\,, \nn\\
d_2&=&2\,\omega[\epsilon_2-|\vec{k}_2 |(\cos{\theta}\,
\cos{(\eta+\phi)}+\cos{\varphi}\sin{\theta}\,\sin{(\eta+\phi)})]\,,
\nn\\
k_2\cdot p_1&=&\epsilon_2\,E-|\vec
p||\vec{k}_2|(\cos{\eta}\cos{\theta}+
\cos{\varphi}\sin{\eta}\sin{\theta})\,.  
\label{eq:47}
\end{eqnarray}
In turn, the respective
trigonometric functions of angles are expressed through the photon
energy and the variable $y,$ as:
\begin{eqnarray}
 \cos{\eta}&=&\frac{|\vec {p}|^2-\omega(E-y)}{|\vec p||\vec{p}-\vec{k}|}\,,
\ \ \cos{(\eta+\phi)}=\frac{E-\omega-y}{|\vec{p}-\vec{k}|}\,, \nn\\
\cos{\theta}&=&\frac{(\epsilon_2-m)(E+m)+
\omega(y+m-\epsilon_2)}{|\vec{p}-\vec{k}|}\,, \ \ \sin{\theta}\,, \
\sin{\eta}\,, \ \sin{(\eta+\phi)}\geq 0\,, \nn\\
|\vec{p}-\vec{k}|&=&\sqrt{|\vec
p|^2+\omega(2y-2E+\omega)}\,. 
\label{eq:47}
\end{eqnarray}
The functions $W_1$ and $W_2$
depend on the azimuthal angle $\varphi$, and, in order  to perform
the integration over this variable in the r.h.s. of Eq.\,(\ref{eq:45}),  one
needs to use a specific expressions for the form factors entering
these functions. Further we concentrate on small values of the
squared momentum transfer as compared with the deuteron mass, where
the form factors can be expanded in a series in term of powers of
$q_1^2.$ In the calculations we keep the terms of the order of
$1\,,$  $q_1^2$, and $q_1^4$ in the quantity
$$-W_1(q_1^2)\,S_1+\frac{W_2(q_1^2)}{M^2}\,S_2$$
which enters the differential cross section.

The integration in the r.h.s. of Eq.\,(\ref{eq:45}) over the $\varphi$ and $y$
variables is performed analytically. The result for both
$C_1(\omega)$ and $C_2(\omega)$ is very cumbersome, and it will be
published elsewhere. In the limit $\omega\to 0$ the function
$C_1(\omega)$ is regular, and the function $C_2(\omega)$ has an
infrared behavior. We extract the regular part $C_{2R}(\omega)$ and
the infrared contribution $C_{2I}(\omega)$ by a simple subtraction
procedure,according to
\begin{equation}\label{eq:49}
C_2(\omega)=\big[C_2(\omega)-C_2(\omega\to 0)\big]+C_2(\omega\to
0)=C_{2R}(\omega)+C_{2I}(\omega)\,, \ \
C_{2I}(\omega)\sim\frac{1}{\omega}
\left[\frac{\epsilon_2}{|\vec{k}_2|}\ln\frac{\epsilon_2+|\vec{k}_2|}{m}-1\right]\,.
\end{equation}
The infrared contribution is combined with the correction due to
soft and virtual photon emission. This results in the substitution 
$\bar\omega\to\omega_s$ in the expression for $\delta_0$, see Eq. (\ref{eq:32}). The integration of the regular part $C_{2R}(\omega)$ over
$\omega$ (we can chose an arbitrary small value as the lower limit)
as well the whole contribution of the region 1, $C_1(\omega),$ is
performed numerically.

%%%%%%%%%%%%%%%%%%%%%%%%%%
\section{Numerical estimations and discussion}

%%%%%%%%%%%%%%%%%%%%%%%%%%

In this section, the conditions for the experimental uncertainties are set to $\Delta E=0.02(E-\epsilon_2)$ and the $t_{20}-$ parametrization
of the deuteron form factors is taken as below, if no other choice is specified.

In our calculation we use four different parameterizations of the deuteron form factors, and
since the four-momentum transfer squared is rather small in this reaction, we can approximate these form factors by a Taylor series expansion
with a good accuracy. On the Born level and when calculating the virtual corrections we can use also unexpanded expressions, but, in order to perform the analytical integrations in Eq.\,(\ref{eq:46}), we have to expand the differential cross section keeping terms up to $q_1^4$ in $W_1(q_1^2)$ and $W_2(q_1^2)$.

So, we use the expansion over the variable $q^2$ of the following four form factor parameterizations.

\underline{By means of the radii (labeled as "{\it rad}\,")}. In this approach we expand the quantity $\cal{D},$ which is defined by Eq.\,(\ref{eq:18}), including terms up to $q^4$ and
we use the expansion of the form factors taking into account only the mean square charge and magnetic radii from Ref. \cite{S01}.
\begin{equation}
\label{eq:50}
\frac{G_{C,M}(q^2)}{G_{E,M}(0)}=1+\frac{1}{6}q^2r^2_{C,M}+O(q^4)\,, \ \ G_Q(q^2)=G_Q(0)\,, \ r_C=2.130 {\mbox{ fm}\,,}\  \ r_M=2.072 \ {\mbox {fm}}\,.
\end{equation}

\underline{"Two-component model for the deuteron electromagnetic
structure" \cite{TGA06} (labeled as "{\it m}")}. In this approach the deuteron form factors are saturated by contribution of the isoscalar
vector mesons, $\omega$ and $\phi.$ In this case one can write:
\begin{equation}\label{eq:51}
G_i(Q^2)=N_i g_i(Q^2) F_i(Q^2),~i=C,\,M,\,Q
\end{equation}
with:
$$F_i(Q^2)= 1-\alpha_i-\beta_i+
\alpha_i\displaystyle\frac{m_{\omega}^2}{m_{\omega}^2+Q^2}
+\beta_i\displaystyle\frac{m_{\phi}^2}{m_{\phi}^2+Q^2}\,, $$
 where $m_{\omega}$ ($m_{\phi}$) is the mass of the $\omega$
($\phi$)-meson. Note that the $Q^2$ dependence of $F_i(Q^2)$ is
parameterized in such form that $F_i(0)=1$, for any values of the
free parameters $\alpha_i$ and $\beta_i$, which are real numbers.

The terms $g_i(Q^2)$ are written as functions of two  real parameters,
$\gamma_i$ and $\delta_i$, generally different for each
form factor:
\begin{equation}\label{eq:52}
 g_i(Q^2)=1/\left [1+\gamma_i\,{Q^2}\right ]^{\delta_i},
\end{equation}
and $N_i$ is the normalization of the $i$-th form factor at $Q^2=0$:
$$N_C=G_C(0)=1, \ N_Q=G_Q(0)=M^2{\cal Q}_d=25.83, \ N_M=G_M(0)=\displaystyle\frac{M}{m_N}\mu_d=1.714,$$
where ${\cal Q}_d$, and $\mu_d$ are the quadrupole and the magnetic
moments of the deuteron.

The experimental data for $G_C$ and $G_M$ show the existence of a
zero, for $Q_{0C}^2\simeq 0.7$ GeV$^2$ and $Q_{0M}^2\simeq 2$
GeV$^2$. The requirement of a node gives the following relation
between the parameters $\alpha_i$ and $\beta_i$, $i=C$ and $M$:
\begin{equation}\label{eq:53}
\alpha_i=\displaystyle\frac{m_{\omega}^2+Q_{0i}^2}{Q_{0i}^2}-
\beta_i\displaystyle\frac{m_{\omega}^2+Q_{0i}^2}{m_{\phi}^2+Q_{0i}^2}.
\end{equation}
The expression (\ref{eq:51}) contains four parameters, $\alpha_i$, $\beta_i$,
$\gamma_i$, $\delta_i$, generally different for different form
factors. The values of the best fit parameters are reported in Table \ref{Table:tab1}. The
common parameters are  $\delta=1.04\pm 0.03$, $\gamma=12.1\pm 0.5$,
 corresponding to $\chi2/ndf=1.1.$ In our calculations we used the central values of this parameters.

\begin{table*}
\begin{tabular}{|c|c|c|c|}
\hline\hline
& $\alpha$ & $\beta$& $\chi2/ndf$  \\
\hline\hline
$G_C$  &$5.75\pm 0.07$ & $-5.11\pm 0.09$ & $0.9$\\
\hline
$G_Q$&$4.21\pm 0.05$ & $-3.41\pm 0.07$ & $0.9$\\
\hline
$G_M$ & $3.77\pm 0.04$& $-2.86\pm 0.05$ & $1.6$\\
\hline\hline
\end{tabular}
\caption{ \label{Table:tab1}Parameters $\alpha$ and $\beta$ for the three
deuteron electromagnetic form factors, from the global fit.
The parameters $\delta$ and $\gamma$ are common to all form factors. 
In case of $G_C$ and $G_M$, $\alpha$ is derived from (\ref{eq:53}).}
\end{table*}

\underline{"Deuteron Electromagnetic Form Factors in the
Transition Region Between  Nucleon-Meson and "}

\underline{Quark-Gluon Pictures \cite{KS94} (labeled as "{\it k}")}. In this approach, 
form factors are consistent with the results from popular NN-potentials at low energies ($Q^2 \ll$ 1(GeV/c)$^2$), but, at the
same time, they provide the right asymptotic behavior following from quark counting rules, at high
energies ($Q^2 \gg$ 1(GeV/c)$^2$). The explicit expressions of 
the deuteron form factors are:
\begin{eqnarray}
G_C&=&\frac{G^2(Q^2)}{(2\tau
+1)}[(1-\frac{2}{3}\tau)A+\frac{8}{3}\sqrt{2\tau}B+\frac{2}{3}(2\tau
-1)C], \ G(Q^2)=\Big(1+\frac{Q^2}{4\,\delta^2}\Big)^{-2},
\nn\\
G_M&=&\frac{G^2(Q^2)}{(2\tau +1)}[2A+\frac{2(2\tau
-1)}{\sqrt{2\tau}}B-2C], \ \ G_Q=\frac{G^2(Q^2)}{(2\tau +1)}[-A+\sqrt{\frac{2}{\tau}}B-\frac{\tau +1}{\tau}C],
\label{eq:54}
\end{eqnarray}
where $\delta$ is some parameter of order of the nucleon mass. The
functions $A$, $B$, and $C$ have the following parametrization:
$$
A=\sum_i^n \frac{a_i}{\alpha_i^2+Q^2}, \ \
B=Q\sum_i^n \frac{b_i}{\beta_i^2+Q^2}, \ \
C=Q^2\sum_i^n \frac{c_i}{\gamma_i^2+Q^2},
$$
where $(a_i, \alpha_i), (b_i, \beta_i), (c_i, \gamma_i)$ are 
fitting parameters. From the quark counting rules it follows that
the fall-off behavior of these amplitudes at high $Q^2$ is
$$A\sim Q^{-2}, \ \ B\sim Q^{-3}, \ \ C\sim Q^{-4},$$
which, together with the requirement of a correct static normalization, 
impose the set of constraints on $(a_i), (b_i), (c_i)$: 
\be\label{eq:55}
\sum_i^n \frac{a_i}{\alpha_i^2}=1, \ \ \sum_i^n b_i=0, \ \ \sum_i^n \frac{b_i}{\beta_i^2}=\frac{2-\mu_d}{2\sqrt{2}M},\
\sum_i^n c_i=0, \ \ \sum_i^n c_i\gamma_i^2=0, \ \ \sum_i^n \frac{c_i}{\gamma_i^2}=\frac{1-\mu_d-{\cal Q}_d}{4M^2}.
\ee
In our calculations we used the following sequence for each group of
these parameters:
$$
\alpha_n^2=2M\mu^{(\alpha)}, \ \
\alpha_i^2=\alpha_1^2+\frac{\alpha_n^2-\alpha_1^2}{n-1}(i-1), \
i=1,...,n
$$
(similarly, for $\beta_i$ and $\gamma_i$), where $\mu^{(\alpha)},
\mu^{(\beta)}$ and $\mu^{(\gamma)}$ have the dimension of energy.
The parameters are listed in Table \ref{Table:table2} for n=4.
\begin{table*}
\begin{tabular}{||c|c|c|c|c||}         \hline
%\multicolumn{5}{|c|}{Table 2} \\ \hline\hline
$\backslash i$ & 1 &  2 &   3 &   4  \\ \hline
$a_{i}\ fm^{-2}$ & $2.4818$ & $-10.850$
& $6.4416$ & see (\ref{eq:55})  \\ \hline
$b_{i}\ fm^{-1}$ & $-1.7654$ & $6.7874$
& see (\ref{eq:55}) & see (\ref{eq:55})  \\ \hline
$c_{i}$ & $-0.053830$ & see (\ref{eq:55}) & see (\ref{eq:55})
& see (\ref{eq:55})  \\ \hline
\multicolumn{3}{||c|}{$\alpha _{1}^{2}=1.8591\ fm^{-2}$} &
\multicolumn{2}{l||}{$\mu^{(\alpha )}=0.58327\ GeV/c$} \\ \hline
\multicolumn{3}{||c|}{$\beta_{1}^2=19.586\ fm^{-2}$} &
\multicolumn{2}{l||}{$\mu^{(\beta  )}=0.1\ GeV/c$} \\ \hline
\multicolumn{3}{||c|}{$\gamma_{1}^2=1.0203\ fm^{-2}$} &
\multicolumn{2}{l||}{$\mu^{(\gamma )}=0.17338\ GeV/c$} \\ \hline
\multicolumn{5}{||c||}{$\delta =0.89852\ GeV/c$}\\ \hline
\end{tabular} 
\caption{Parameters corresponding to Eq. (\protect\ref{eq:55}) for n=4.}
\label{Table:table2}
\end{table*}

\underline{"Jefferson $t_{20}$ collaboration" (labeled as {\small\,${20}$)}} \cite{A2000}.
The three deuteron electromagnetic form factors have been determined by fitting directly the all existing measured differential cross
section and polarization observables, according to the following expressions:
\be\label{eq:56}
G_i(Q^2)=G_i(0)\,D_i(Q^2)\,I_i(Q^2)\,, \ \ D_i(Q^2)=1-\frac{Q^2}{Q_i^2}\,, \ I_i(Q^2)=\frac{1}{1+S_i(Q^2)}\,, \ i=C,\,M,\,Q,
\ee
where
$$S_i(Q^2)=\sum_{k=1}^5\,a_i^k\,Q^{2k}\,, \ Q_C^2=17.72\, {\mbox{fm}}^2, \ Q_M^2=54.32\, {\mbox {fm}}^2, \ Q_Q^2=65.61\,{\mbox {fm}}^2\,.$$

The parameters $a_i^k$ have such dimensions of the inverse $Q^2$ powers to quantities $S_i(Q^2)$ were dimensionless.
The values of these parameters appear in Table \ref{Table:table3} with $Q^2$ in fm$^2$ units.

\begin{table*}
\begin{tabular}{||c|c|c|c|c|c|c||}         \hline
%\multicolumn{6}{|c|}{Table 3} \\ \hline\hline
$\backslash k$ & 1 &  2 &   3 &   4 & 5 \\ \hline
$a_C$ & 0.674 & 0.02246 & 0.009806 & -0.0002709 & 0.000003793  \\ \hline
$a_M$ & 0.5804 & 0.08701 & -0.003624 & 0.0003448 & -0.000002818  \\ \hline
$a_Q$ & 0.8796 & -0.5656 & 0.01933 & -0.0006734 & 0.000009438  \\ \hline
\end{tabular} \\
\caption{Parameters corresponding to Eq.(\protect\ref{eq:56}) in fm$^2$ units.}
\label{Table:table3}
\end{table*}

\begin{figure}[t]
\centering
\includegraphics[width=0.5\textwidth]{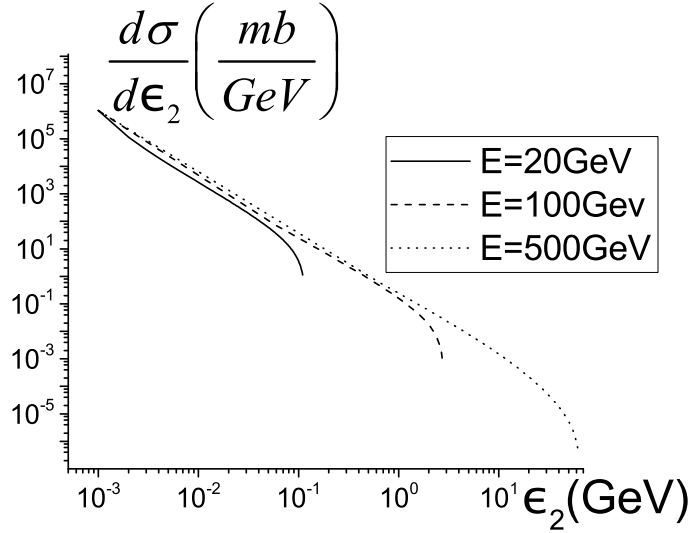}
 \caption{Born differential cross section, defined by Eq. (\ref{eq:17}), is calculated with the $t_{20}$ parametrization of the form factors at different beam energies.}
 \label{Fig:elastde}
\end{figure}

To illustrate the dependence of the recoil electron distribution on the deuteron beam energy, the Born cross section is shown in Fig. \ref{Fig:elastde} for standard $t_{20}$
parametrization at E=20\,GeV, 100\,GeV and 500\,GeV. Here and below, for the beam energy 500\,GeV we limi the recoil electron energy to 10\,GeV,
because for largest values the expansions for the form factors are incorrect.

\begin{figure}[h]
\centering
\includegraphics[width=0.3\textwidth]{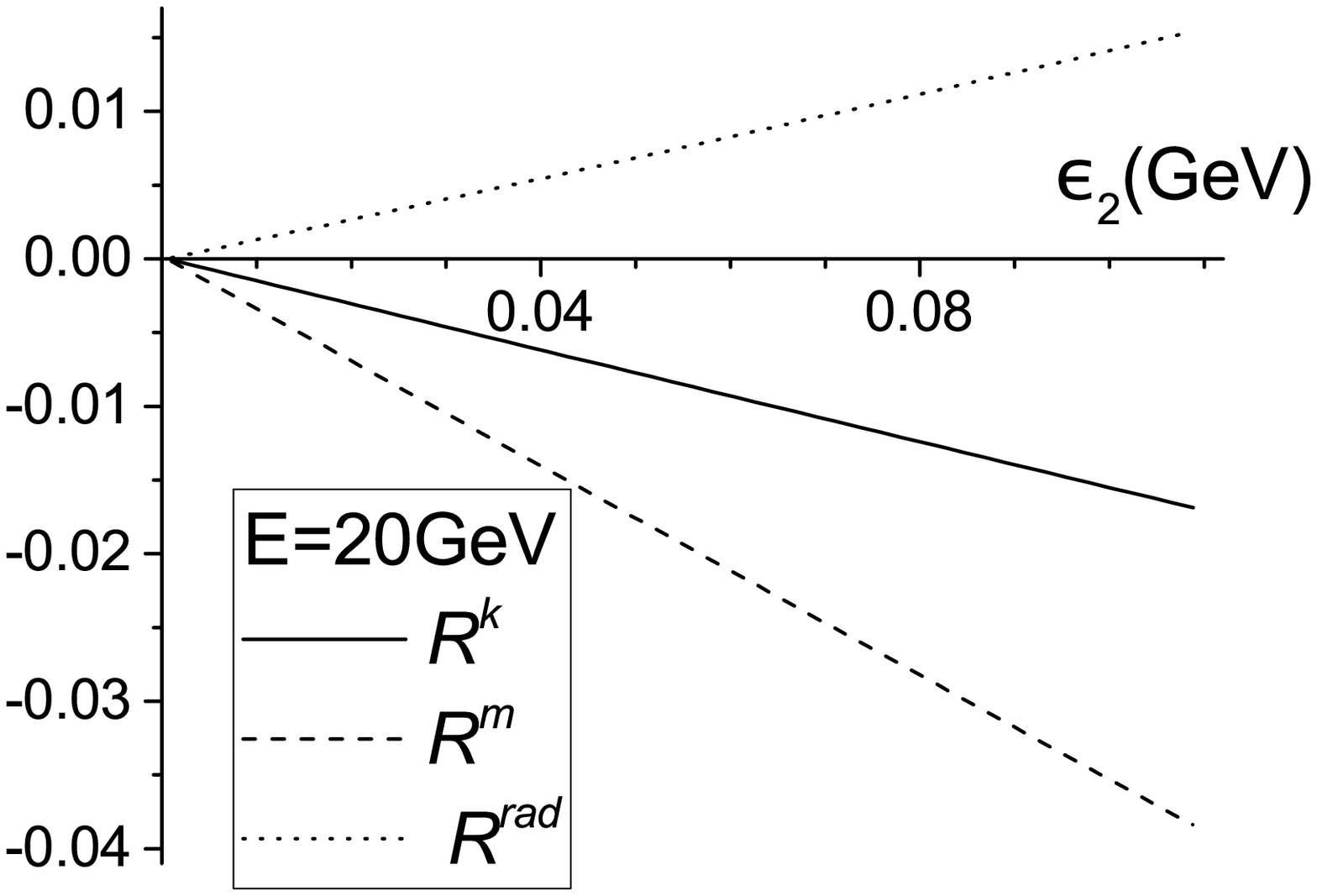}
\includegraphics[width=0.3\textwidth]{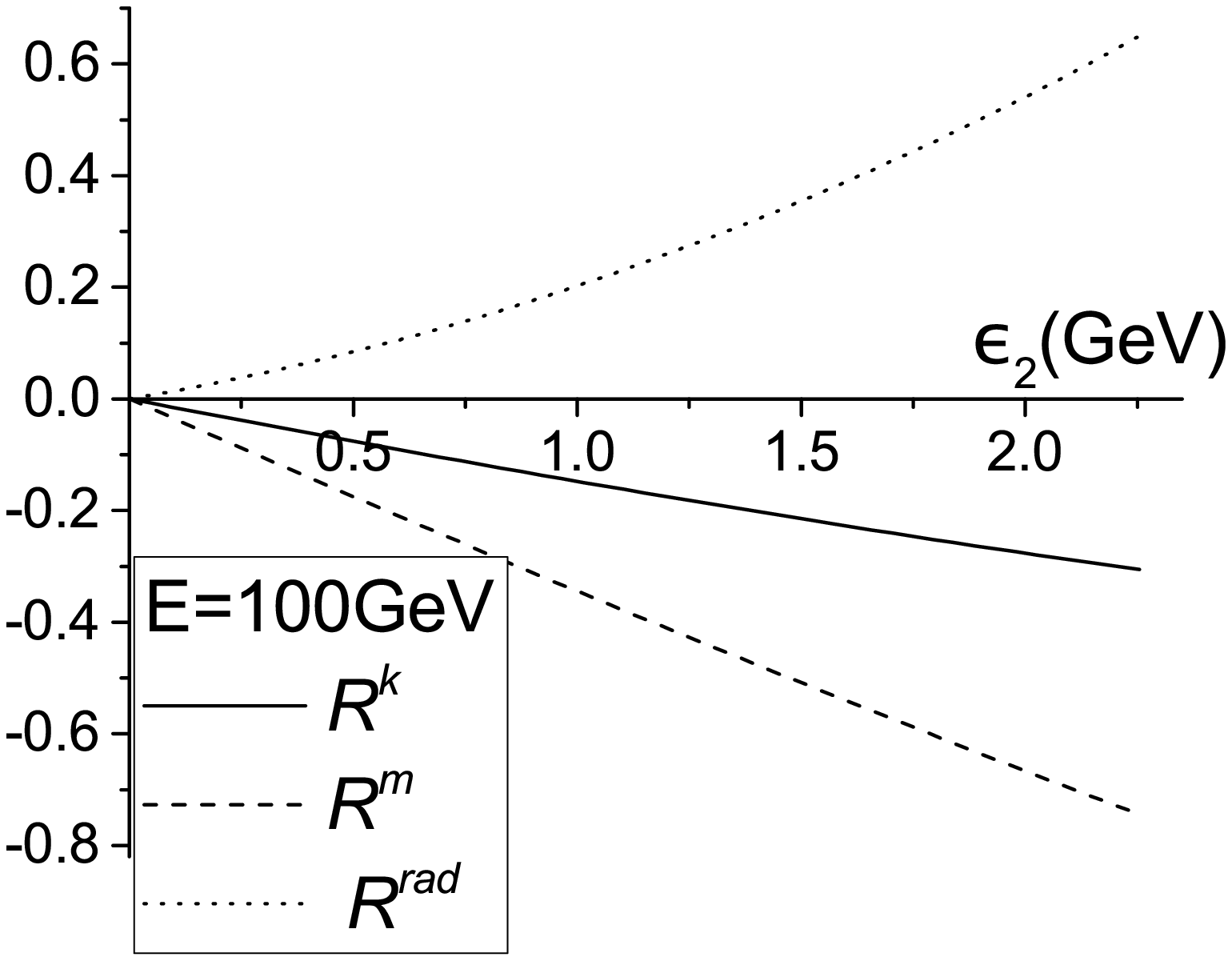}
\includegraphics[width=0.3\textwidth]{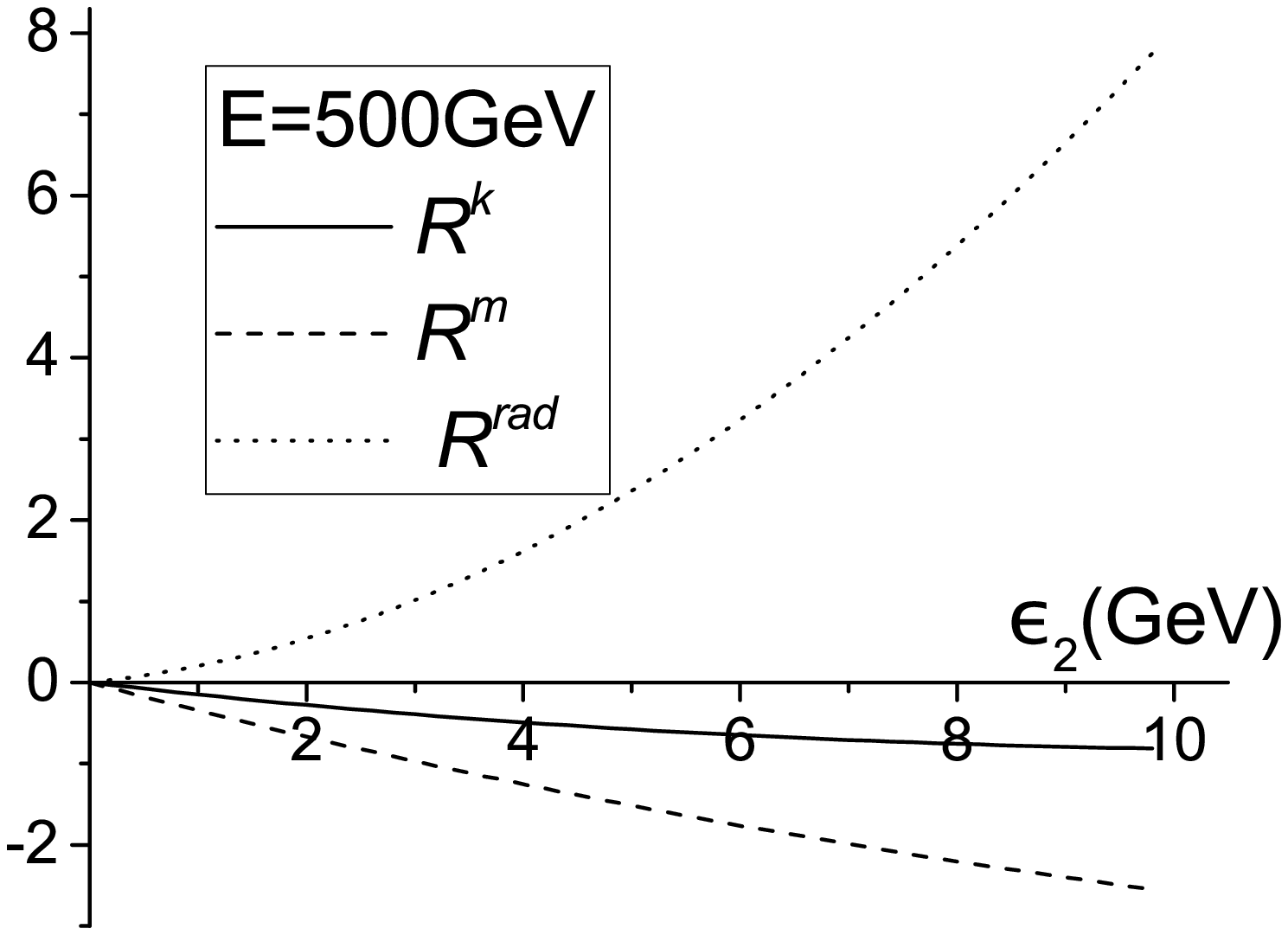}

\vspace{0.5cm}

\includegraphics[width=0.3\textwidth]{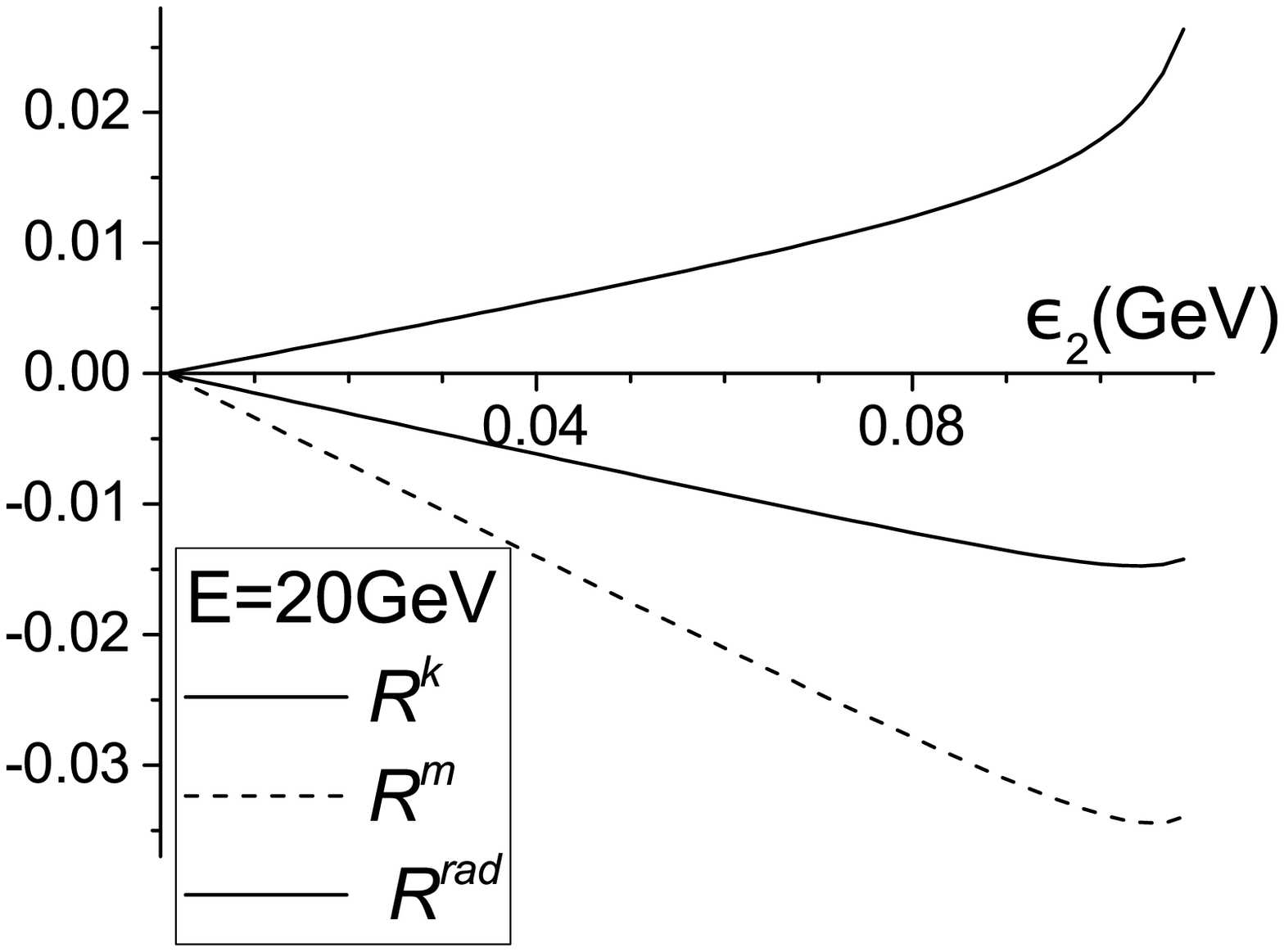}
\includegraphics[width=0.3\textwidth]{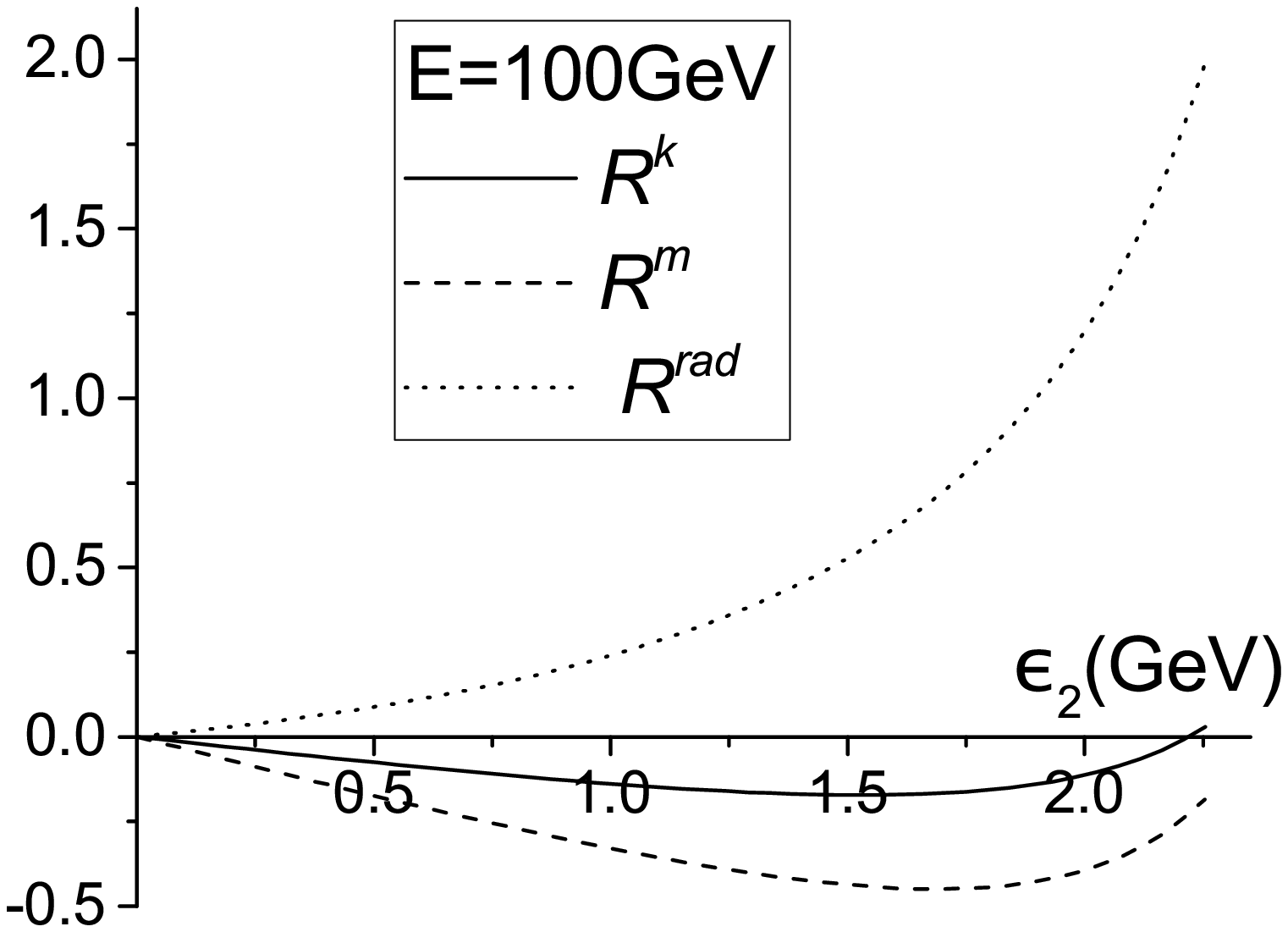}
\includegraphics[width=0.3\textwidth]{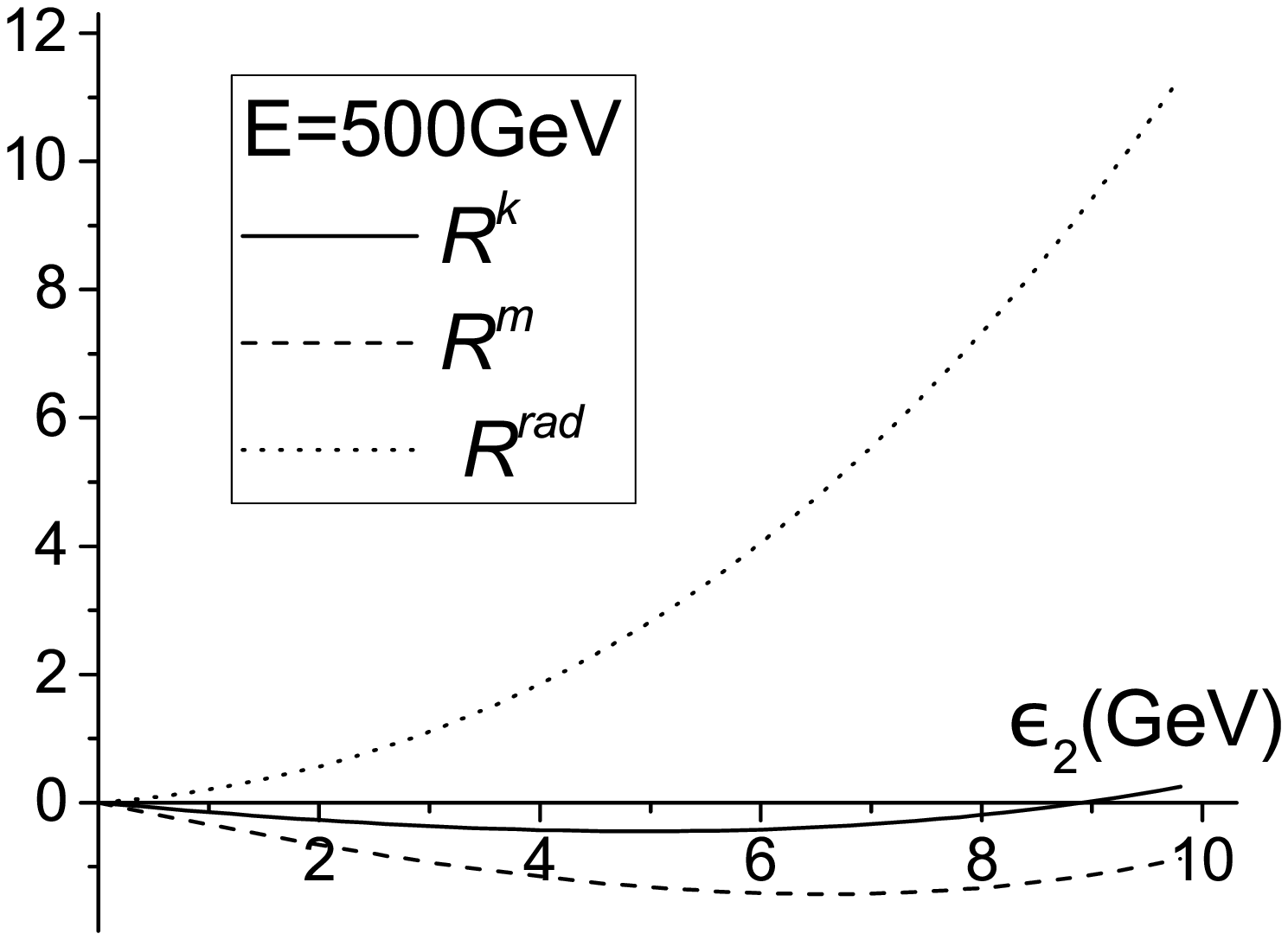}
\caption{Difference of the recoil electron distributions, Eq.~(\ref{eq:57-R}), in percent,  normalized to $d\,\sigma^{20}$, for various  parameterizations of the form factors, at deuteron energies 20 GeV, 100 GeV and 500 GeV. The upper set corresponds to unexpanded form factors and the lower
one to expanded form factors, keeping the terms up to $q^4$.}
 \label{Fig:R}
\end{figure}

The sensitivity of this cross section to different form factor parameterizations is shown in Fig. \ref{Fig:R}, in terms of the quantities (in percentages)
\begin{equation}\label{eq:57-R}
R^k=1-\frac{d\,\sigma^k}{d\,\sigma^{20}}\,, \ R^m=1-\frac{d\,\sigma^m}{d\,\sigma^{20}}\,, \ R^{rad}=1-\frac{d\,\sigma^{rad}}{d\,\sigma^{20}}\,,
\end{equation}
where $d\,\sigma^i$ is the differential cross section (\ref{eq:17}), and {\it i=k,\,m,\,rad,\,\small${20}$} correspond to the above-mentioned deuteron form factors.
As one can see, the sensitivity has a very similar behaviour for the expanded and unexpanded cross sections and increases when  both the deuteron and recoil electron energies increase. However, the differential cross section decreases very quickly when the recoil electron energy increases (see Fig. \ref{Fig:elastde}).

\begin{figure}[t]
\centering
\includegraphics[width=0.31\textwidth]{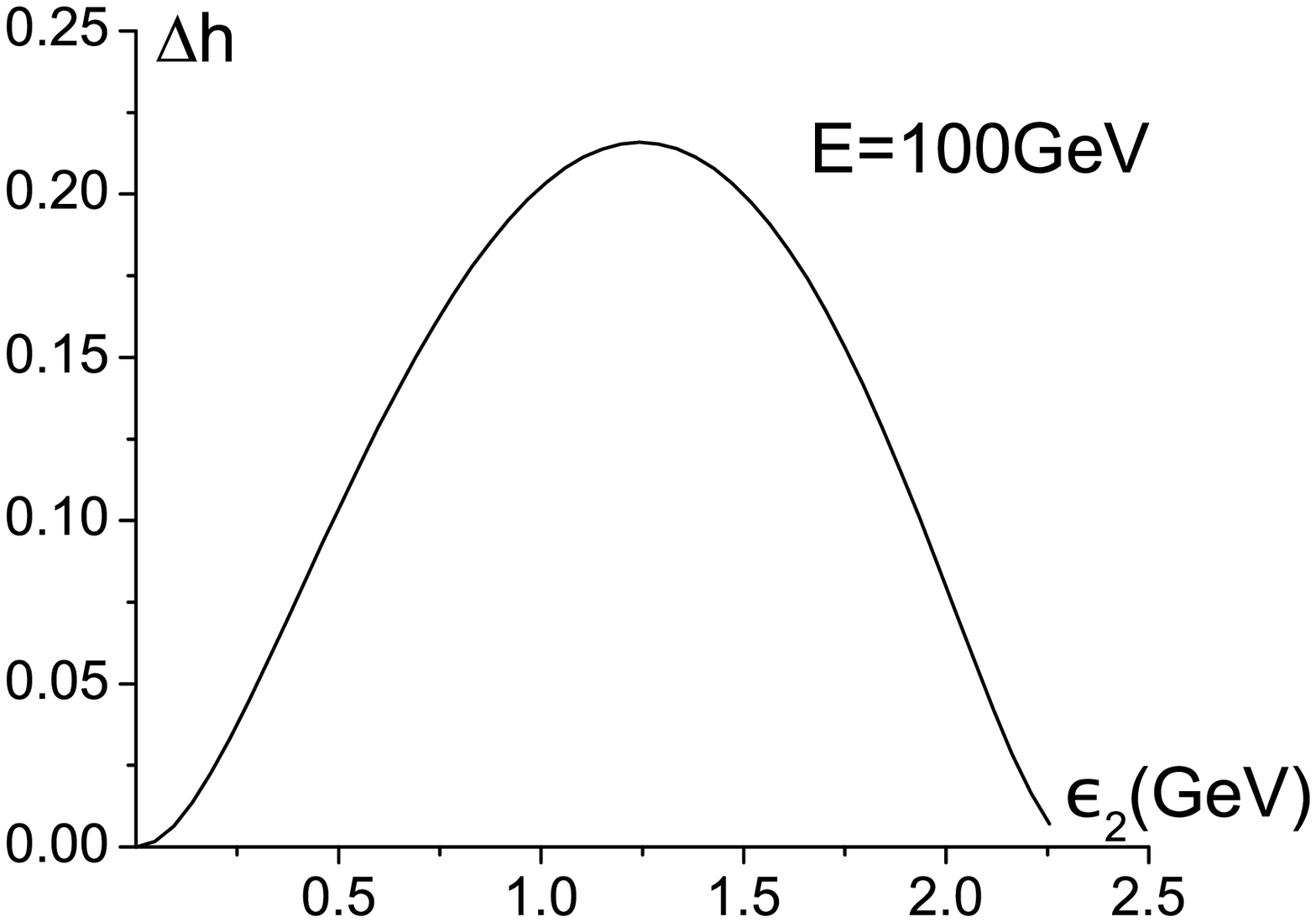}
\includegraphics[width=0.31\textwidth]{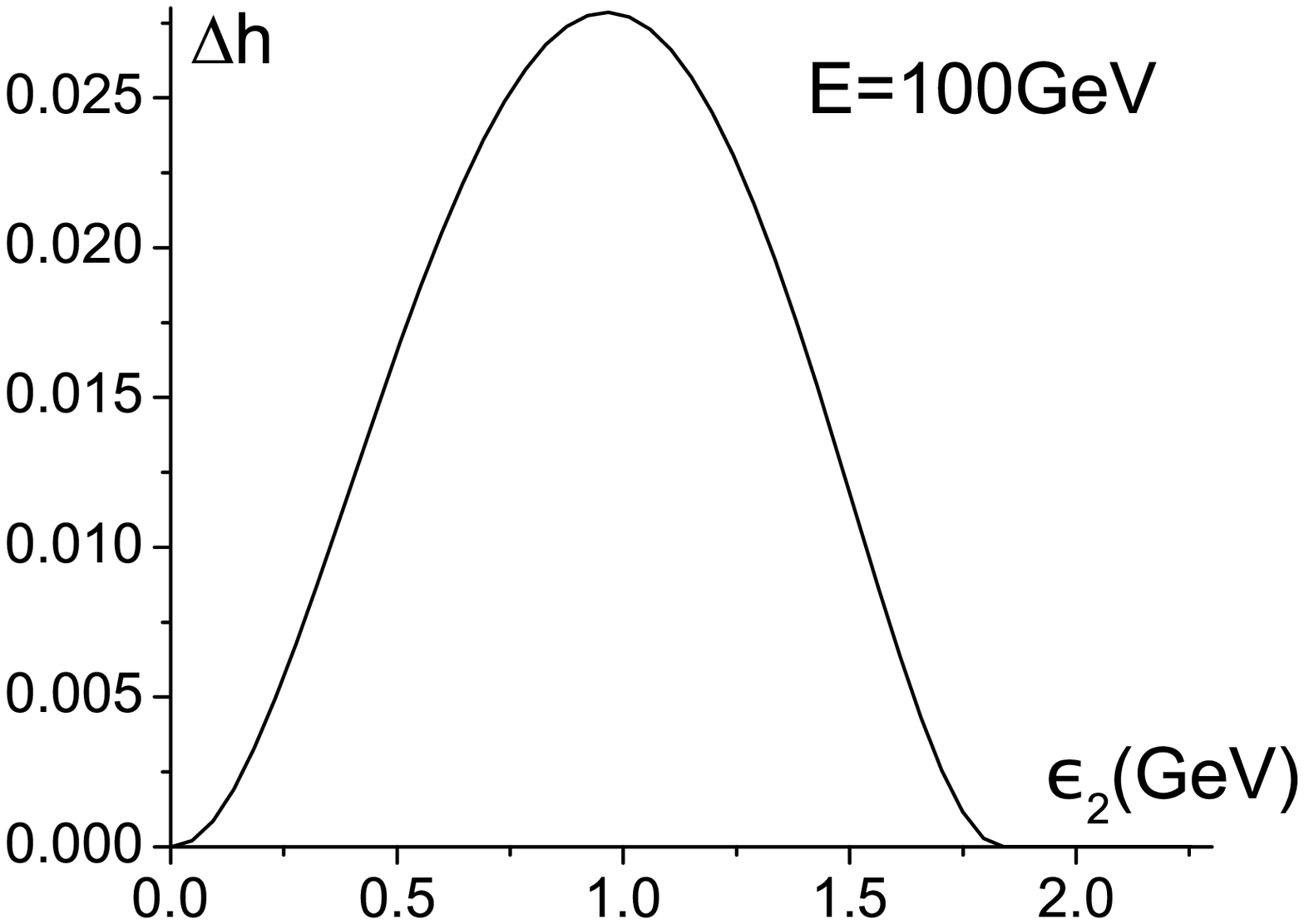}
\includegraphics[width=0.31\textwidth]{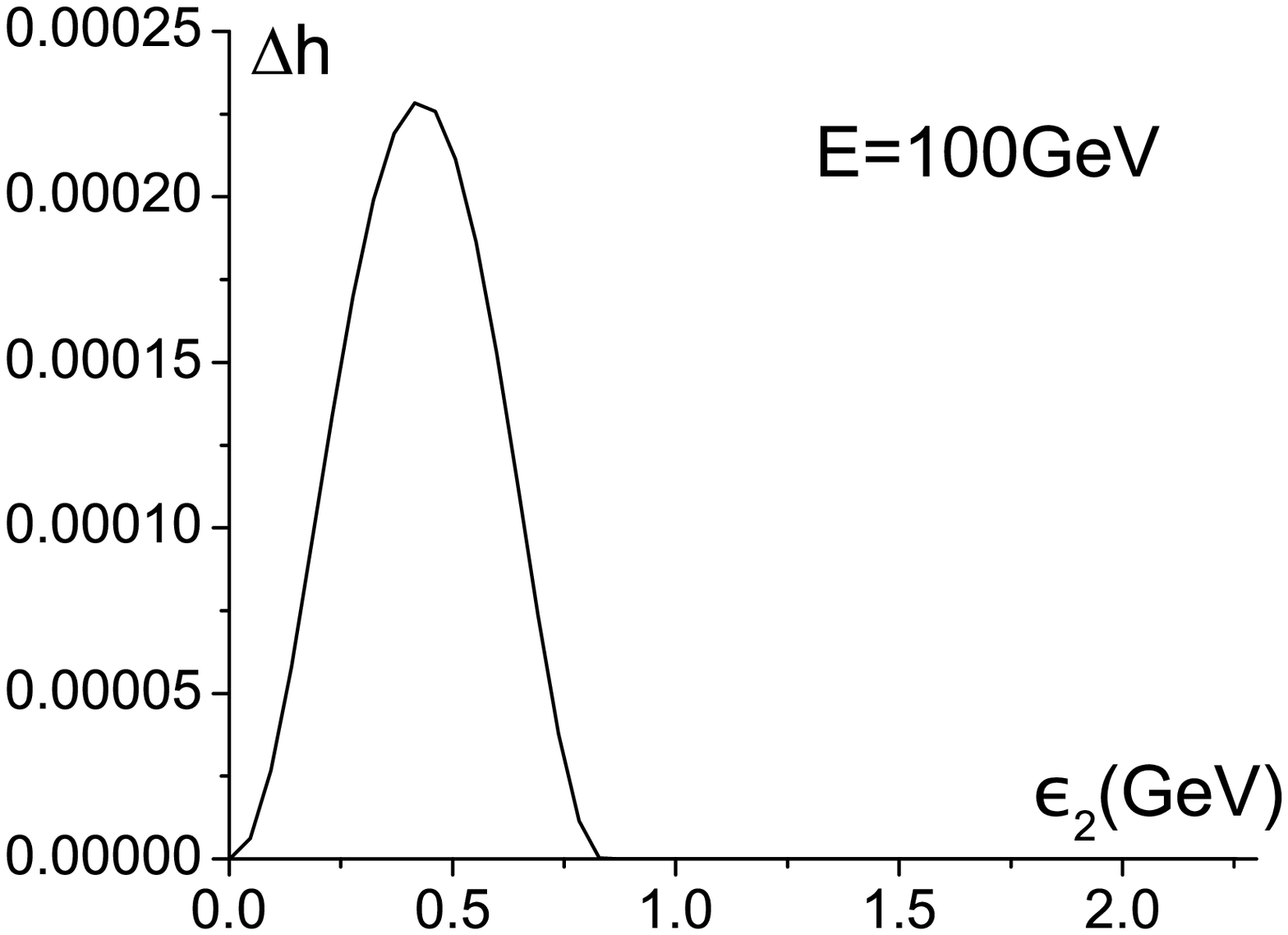}

\vspace{0.5cm}

\includegraphics[width=0.31\textwidth]{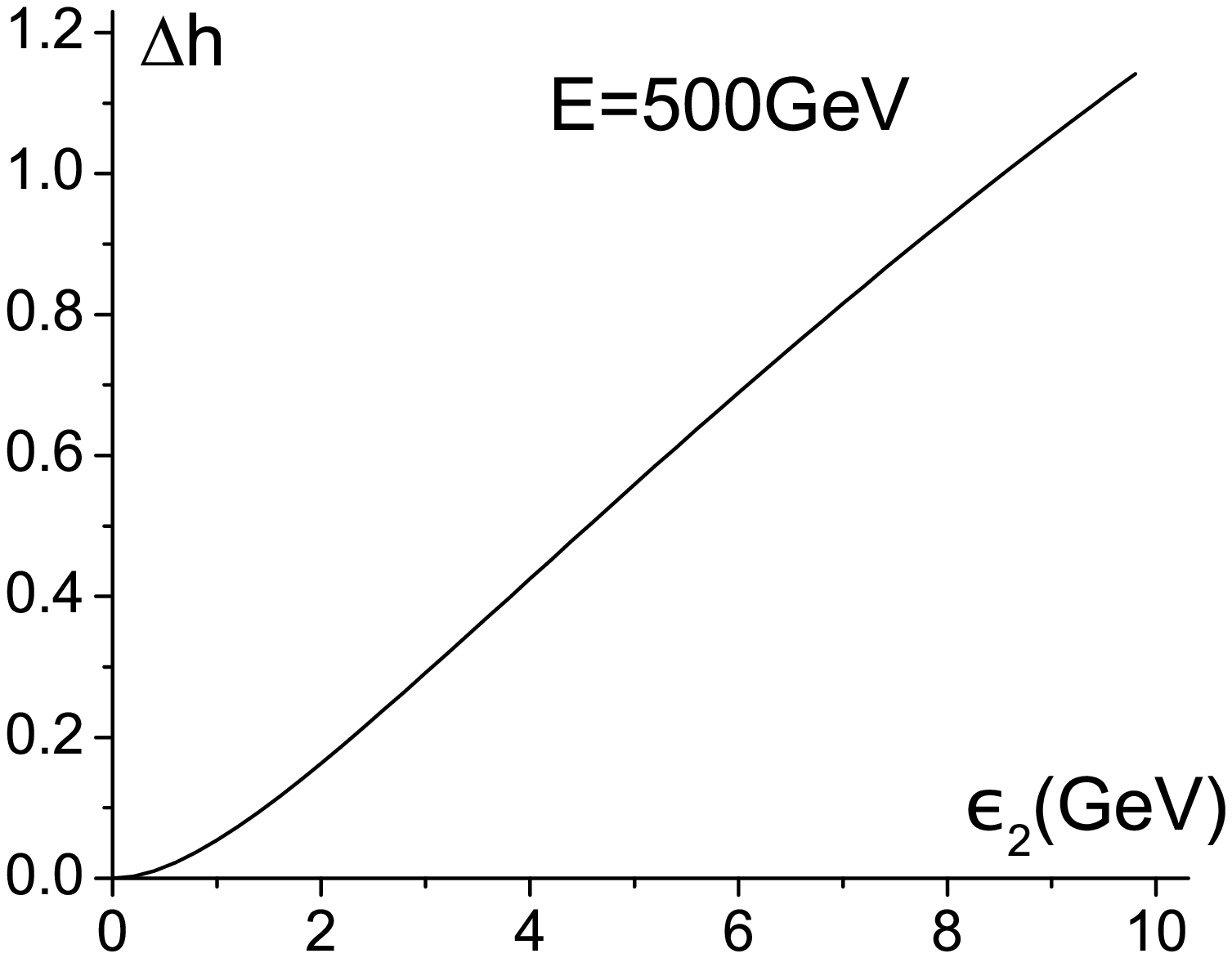}
\includegraphics[width=0.31\textwidth]{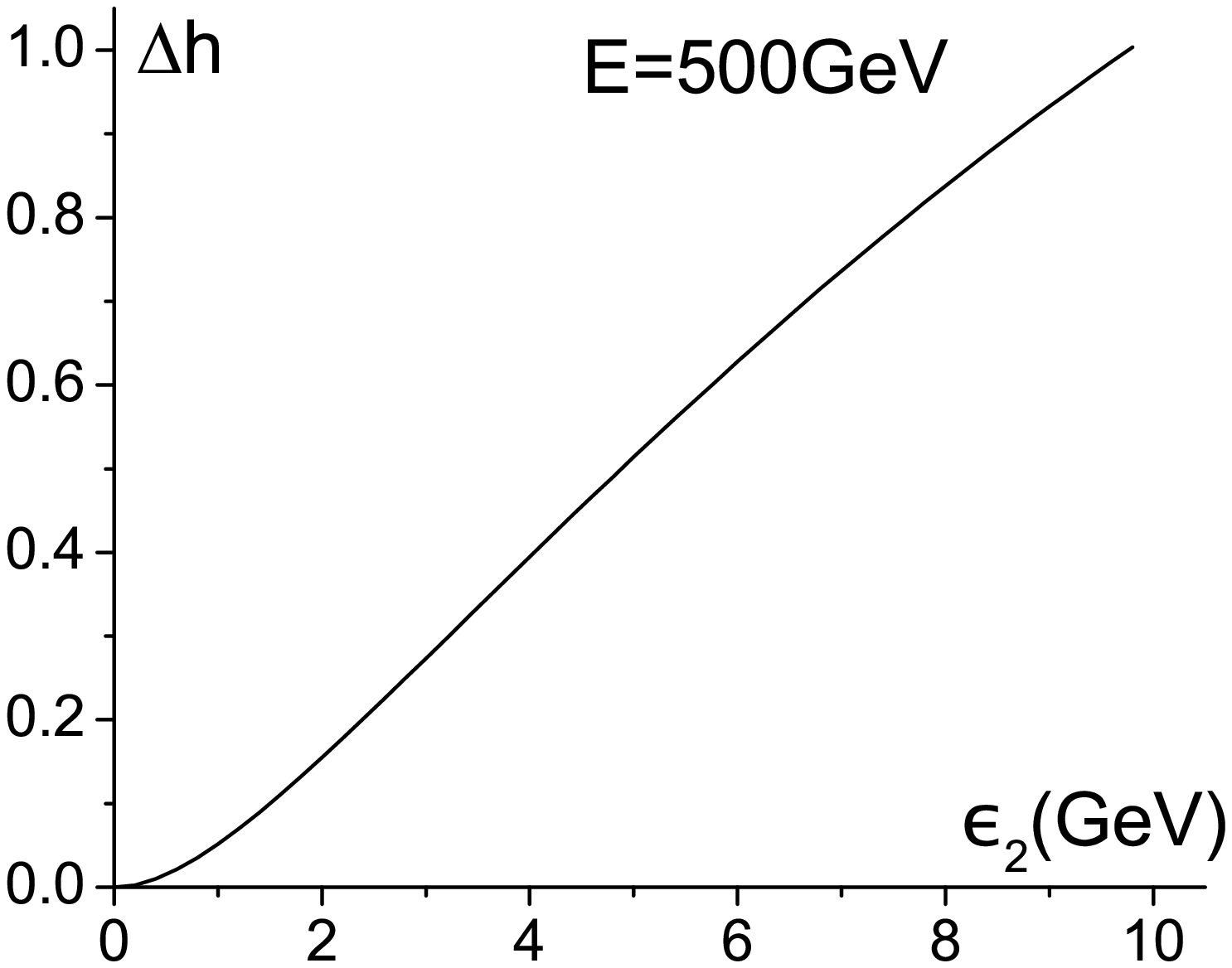}
\includegraphics[width=0.31\textwidth]{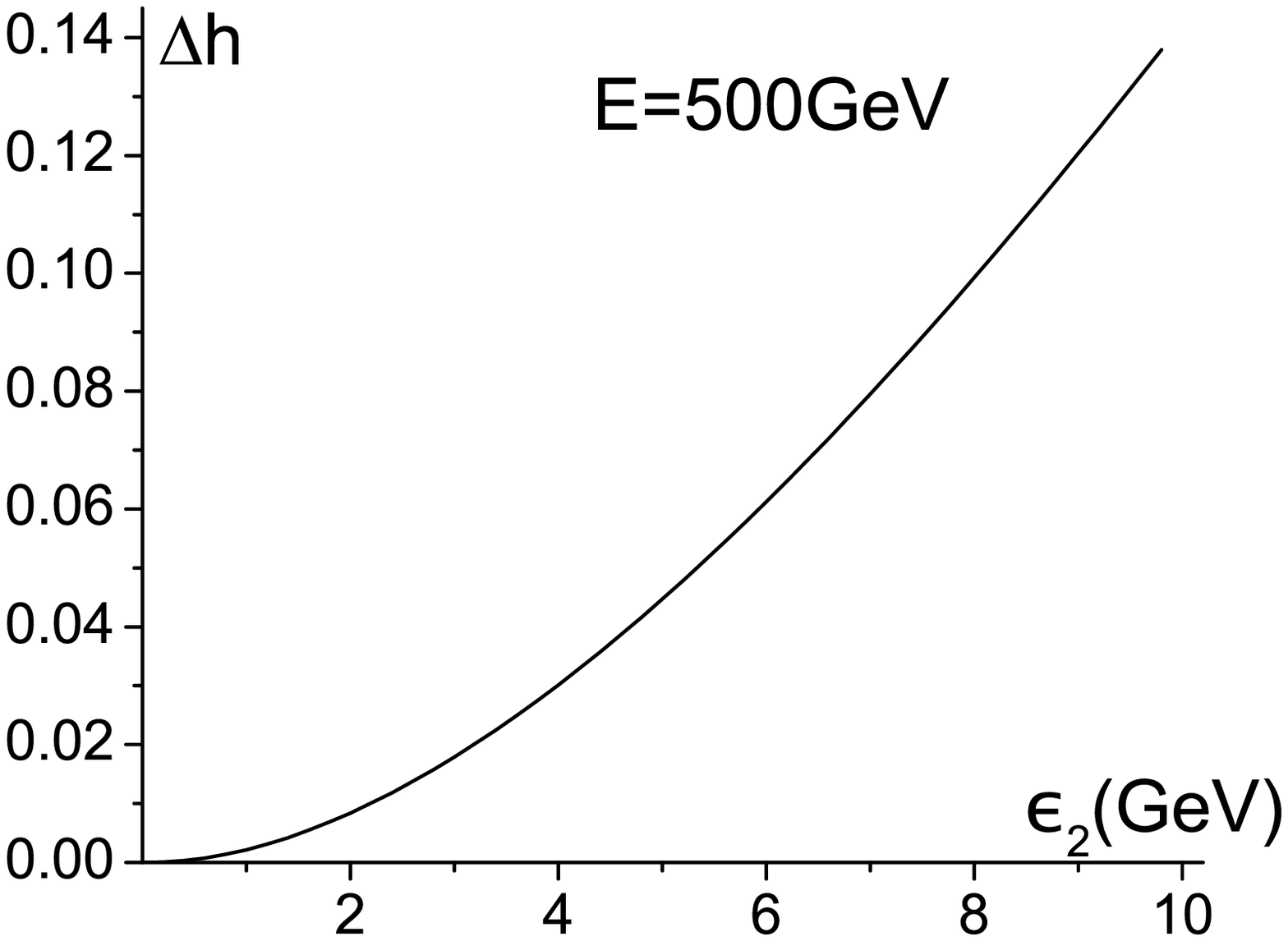}
\caption{The quantity $\Delta h$ (in percent)\, calculated according to Eq.\,(\ref{eq:58-delh}), as a function of the recoil electron energy, at  deuteron energy of 100 GeV  (upper row) and 500 GeV (lower row). The left panels correspond to c=0.005, the middle ones $-$ to c=0.01 and
the right ones $-$ to c=0.02.}
\label{Fig:fig8}
\end{figure}

The hard photon correction depends on the parameter $\Delta E$ due to the contribution of the region 1 in Fig. \ref{Fig:fig5} (left panel). To illustrate this
dependence, we show in Fig. \ref{Fig:fig8} the quantity (in which the contribution of the region 2 is removed)
\begin{equation}\label{eq:58-delh}
\Delta\,h=\frac{d\,\sigma^{(h)}(\Delta E=0.05 (E_2-\epsilon_2))-d\,\sigma^{(h)}(\Delta E=c_i (E_2-\epsilon_2))}{d\,\sigma^{(B)}}\,,
\ \ c_1=0.005,\, c_2=0.01,\, c_3=0.02
\end{equation}
as a function of the recoil electron energy for the $t_{20}$ parametrization. The effect is rather small: on the level $1\%\,(0.1\%)$ for E=500\,(100)\,GeV.
If the deuteron energy is 500\,GeV (lower row) this dependence exhibits the monotonic increase with the recoil electron energy, whereas at
the energy 100\,GeV (upper row) it has maximum and then decreases up to zero. In this zero-point, the recoil-electron energy value is the root of the
equation $c_i(E_2-\epsilon_2)=\omega_{max}(\epsilon_2),$ provided that line $\Delta E$ in Fig.\,4 lies above the curve $\omega_{max}(\epsilon_2).$
At the deuteron energy 500\,GeV this last condition is not fulfilled.

\begin{figure}[t]
\centering
\includegraphics[width=0.3\textwidth]{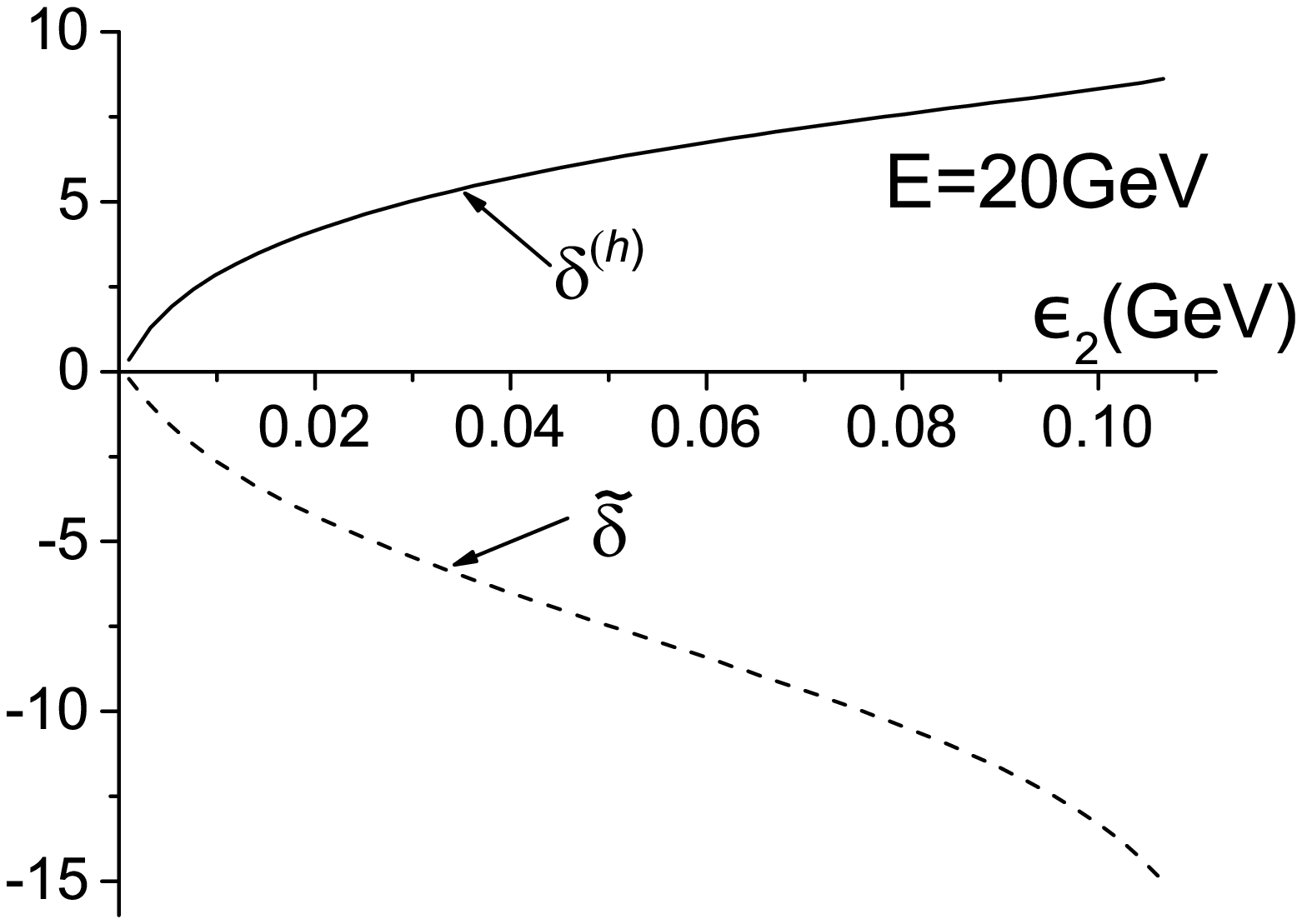}
\includegraphics[width=0.3\textwidth]{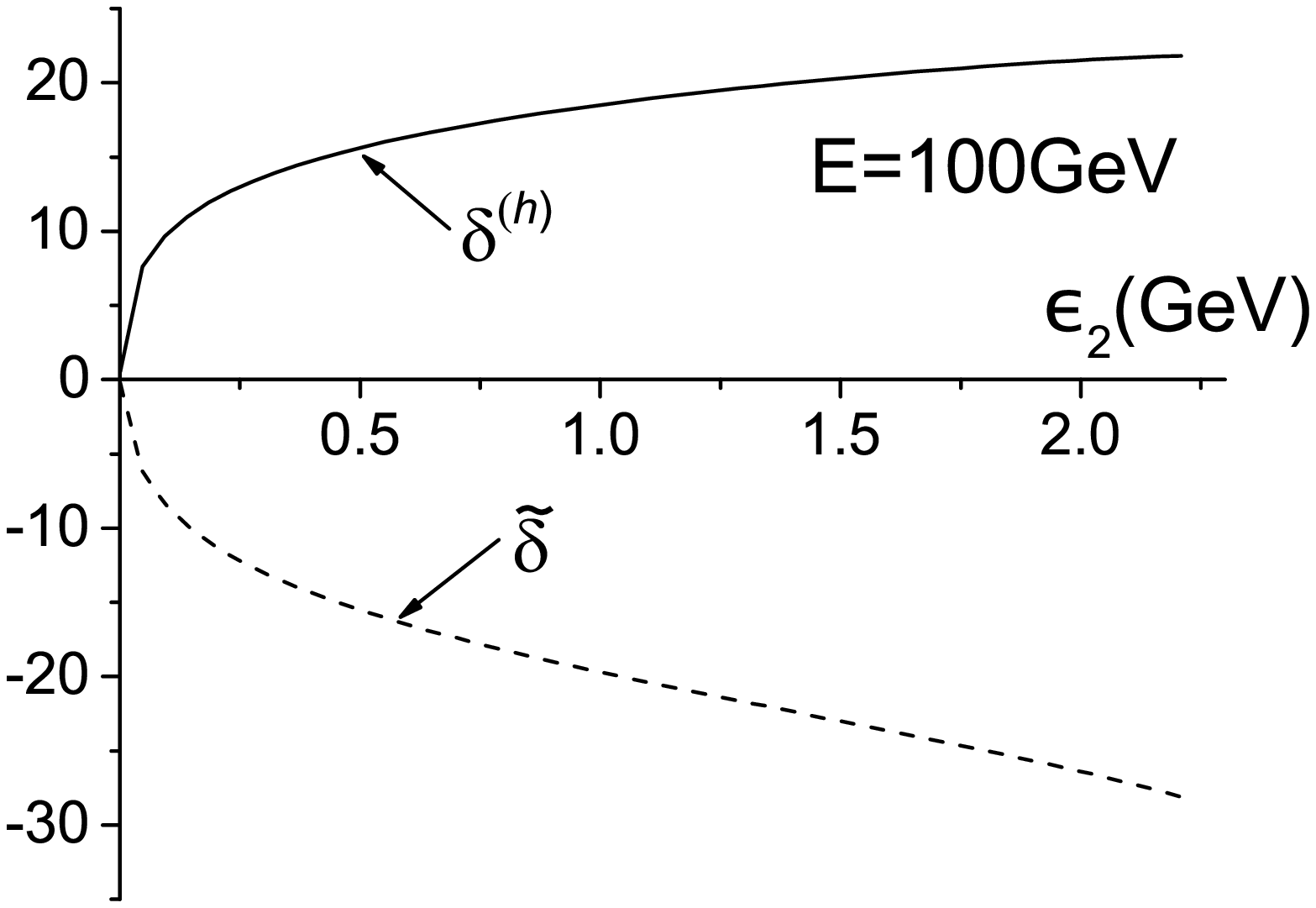}
\includegraphics[width=0.3\textwidth]{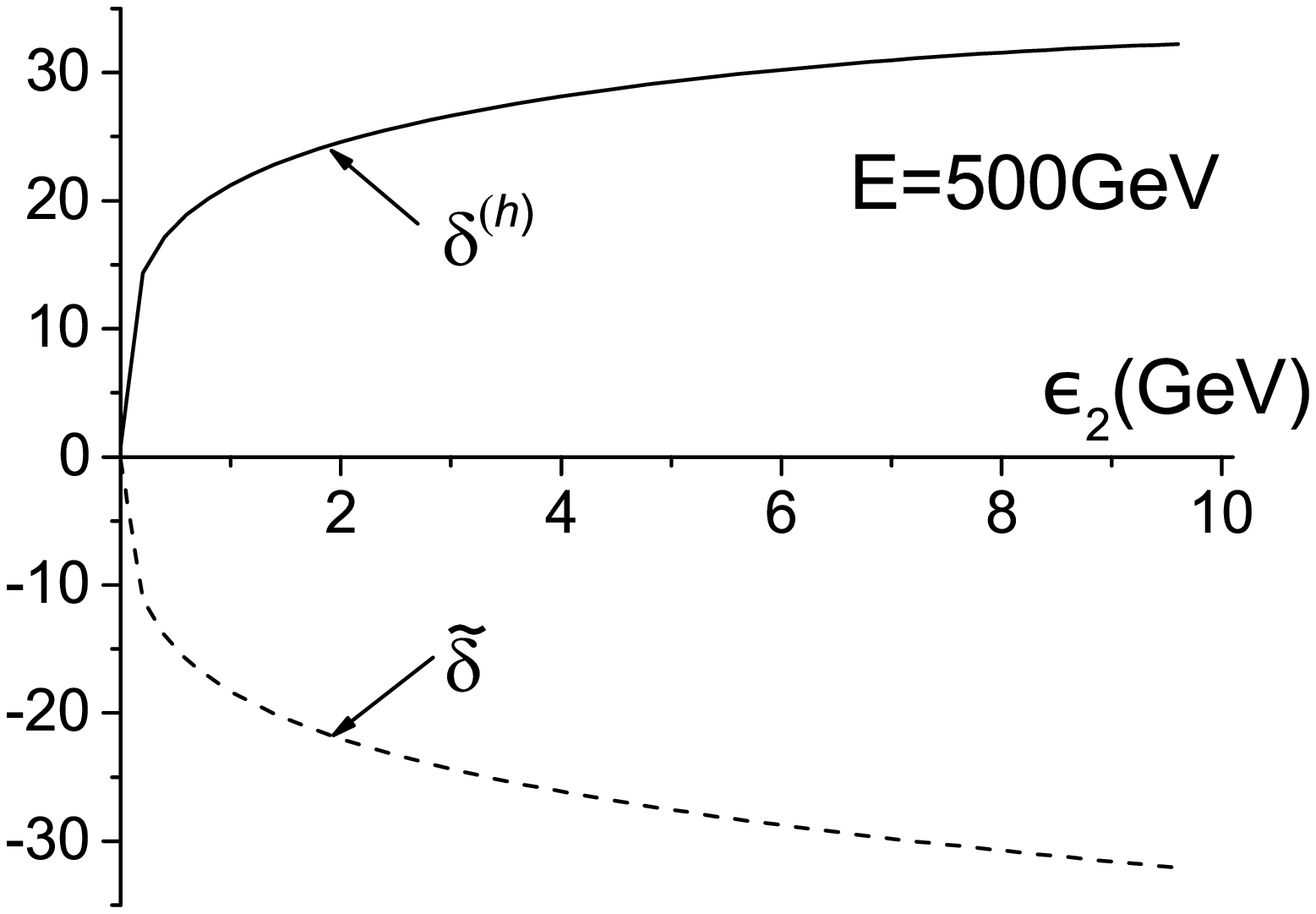}

\vspace{0.6cm}
\includegraphics[width=0.3\textwidth]{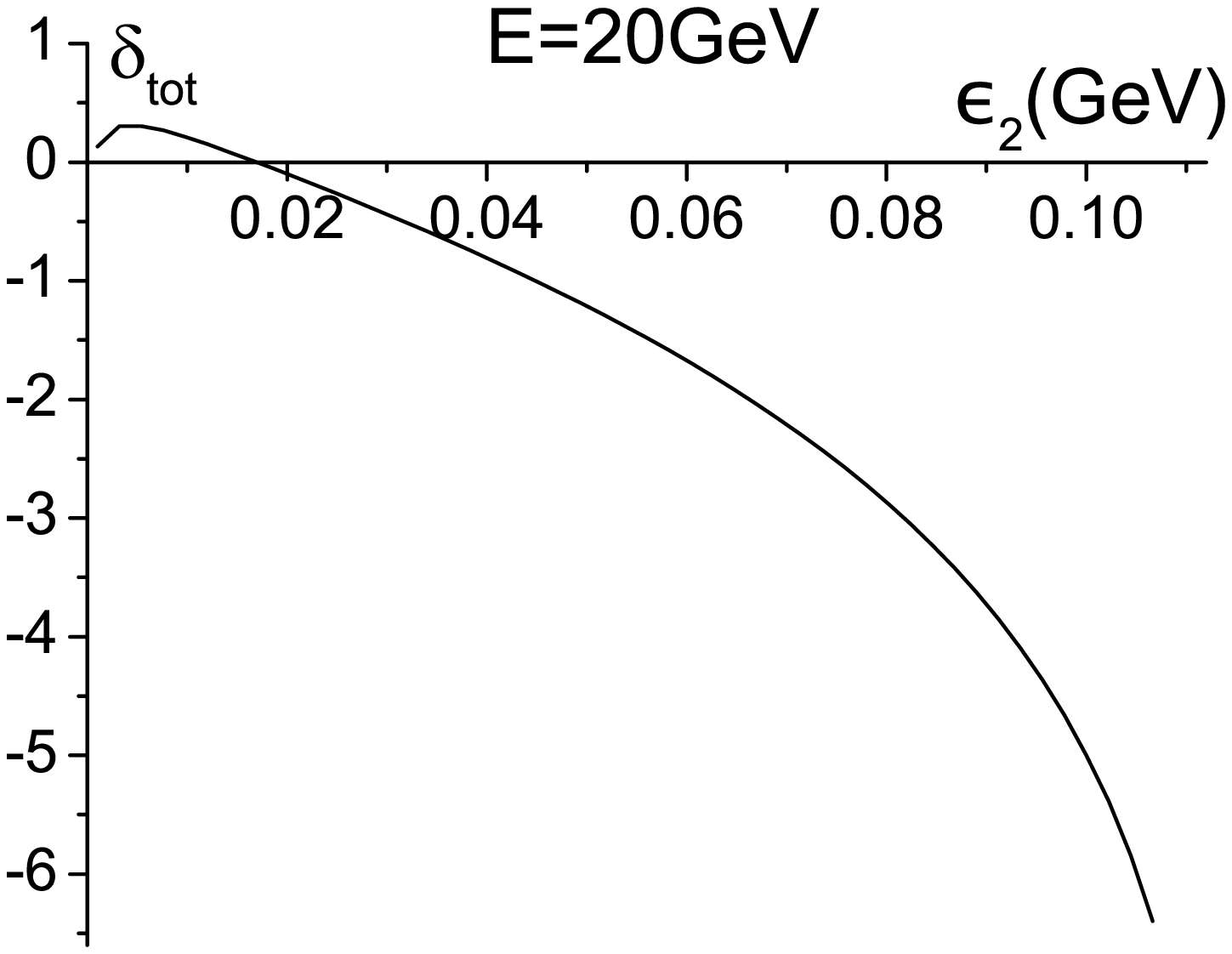}
\includegraphics[width=0.3\textwidth]{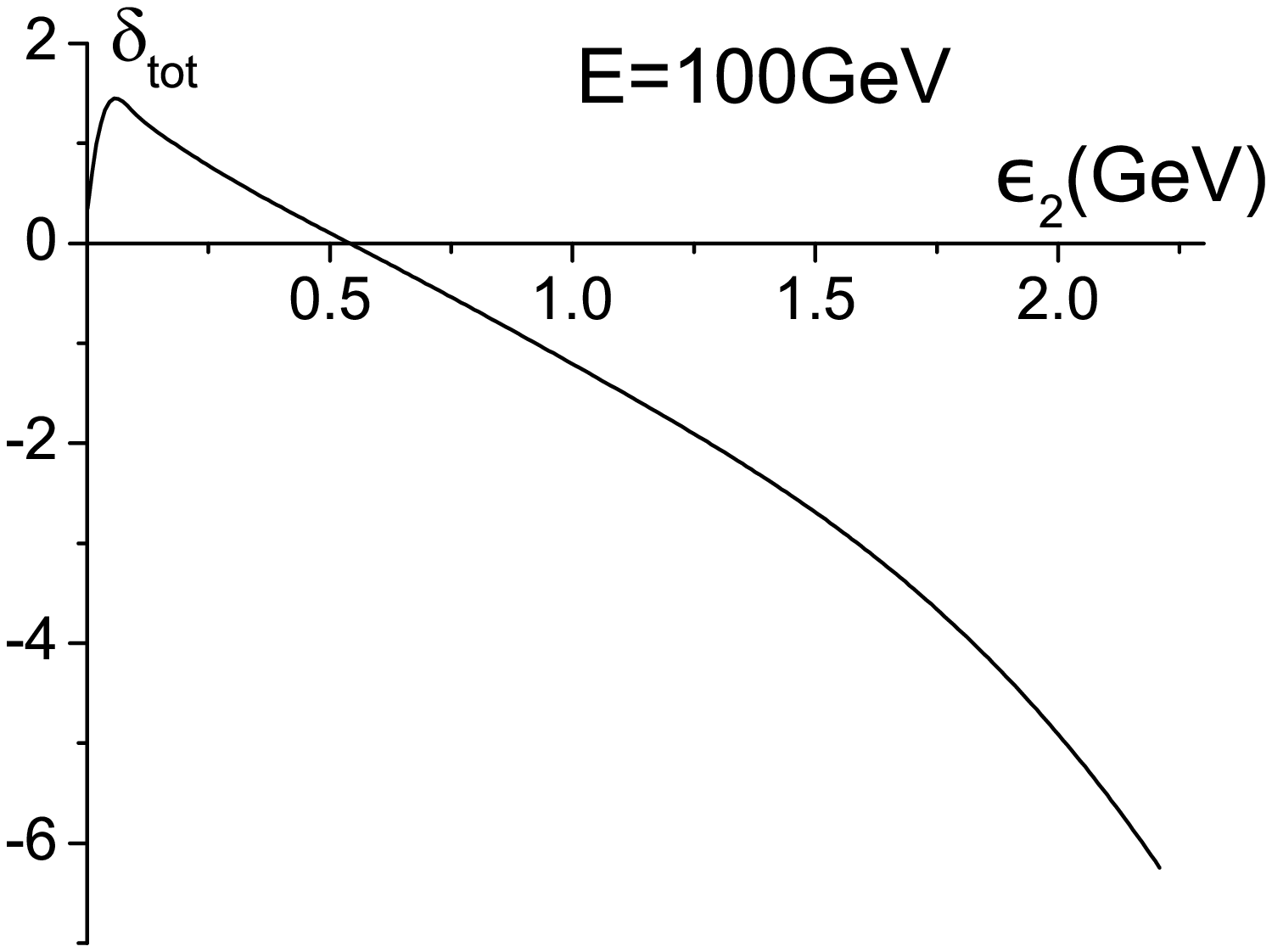}
\includegraphics[width=0.3\textwidth]{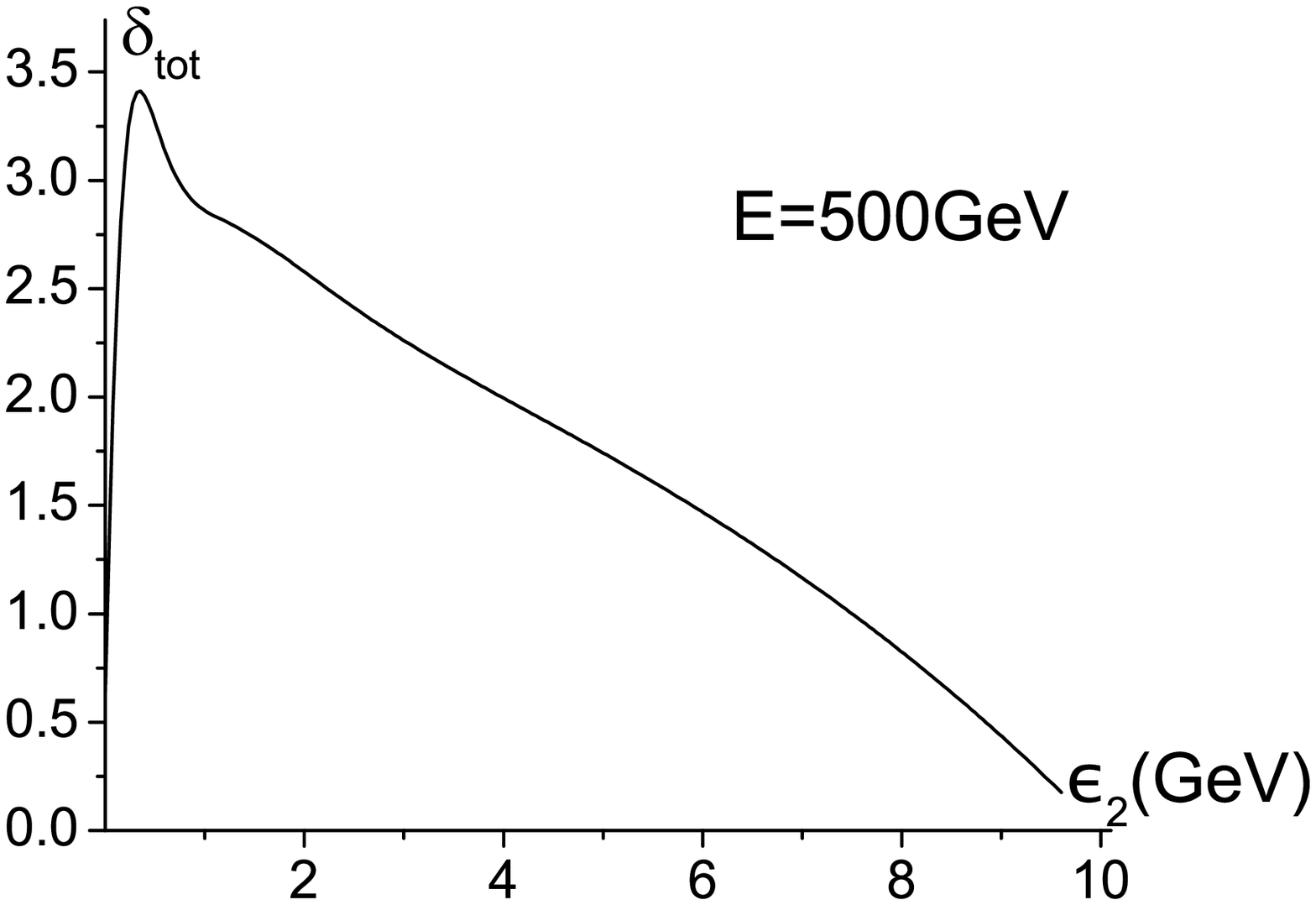}
\caption{(Top) The modified soft and virtual ($\widetilde\delta$)  (dashed line) and hard $(\delta^{(h)})$ (solid line) corrections (in percent) as defined by Eq.\,(\ref{eq:59}). (Bottom)
 The total radiative correction (in percent) calculated for the standard $t_{20}$ fit at $\Delta E =0.02\,(E-\epsilon_2)$, and 20\,GeV(left), 100 GeV (middle) and 500 GeV (right) incident deuteron energy.}
 \label{Fig:fig9}
\end{figure}

In Fig. \ref{Fig:fig9} we present the quantities $\delta^{(h)}$ and $\widetilde{\delta},$
defined as
\ba\label{eq:59}
\delta^{(h)}=\frac{d\,\sigma^{(h)}}{d\,\sigma^{(B)}}-\frac{2\,\alpha}{\pi}\ln\frac{\omega_s}{\bar{\omega}}
\left[\frac{\epsilon_2}{|\vec{k}_2|}\ln\left(\frac{\epsilon_2+|\vec{k}_2|}{m}\right)-1\right ]\,, \nn \\ \widetilde{\delta}=\bar{\delta}+\delta^{(\rm vac)}+\frac{2\,\alpha}{\pi}\ln\frac{\omega_s}{m}
\left[\frac{\epsilon_2}{|\vec{k}_2|}\ln\left(\frac{\epsilon_2+|\vec{k}_2|}{m}\right )-1\right ]\,,
\ea
which we call  "modified hard and soft and virtual corrections", respectively,
as well their sum $\delta_{\rm tot}=\delta^{(h)}+\widetilde{\delta}$ that is the total model-independent first order radiative correction
(the last term in $\tilde{\delta}$ is $\delta_0(\bar{\omega}\to \omega_s))$: 
$$\delta_{\rm tot}=\delta^{(h)}+\tilde\delta=\delta_{0}+\bar{\delta}+\delta^{(\rm vac)}+\frac{d\sigma^{(h)}}{d\sigma^{(B)}}\,.$$
Note, that both modified corrections in Eq. (\ref{eq:59}) are independent on the auxiliary parameter $\bar\omega$ but depend on the physical parameter $\omega_s$ and, therefore, have a physical sense.

To calculate
$\delta_{\rm tot},$ we can write the quantity $(1+\delta_0(\bar{\omega}\to \omega_s))$
using the expression (\ref{eq:32}) or its exponential form defined by (\ref{eq:34}) (with substitution $\bar{\omega}\to \omega_s$). But numerical estimations show that they differ very insignificantly, by a few tenth of the percent, and further we do not use the exponential form.

We see that at small values of the squared momentum transfer (small recoil-electron energy $\epsilon_2$) the total model-independent radiative correction is positive and it decreases (with increase of $\epsilon_2$), reaching zero and becoming negative. The absolute value of the radiative correction does not exceed 6$\%,$ although the strong compensation of the large (up to 30 \%) positive "modified hard" and negative "modified soft and virtual" corrections takes place. Such behavior of the pure QED correction is similar to one derived in Ref.\,\cite{K64}.

If the deuteron form factors are determined independently with high accuracy from other experiments, the measurement of the cross section $d\,\sigma/d\,\epsilon_2$ can be used, in principle, to
measure the model-dependent part of the radiative correction in the considered conditions. This possibility is similar to the one described in
Ref.\,\cite{Abbiendi:2017}
where the authors proposed to determine the hadronic (model-dependent) contribution to the running electromagnetic coupling $\alpha(q^2)$ by a precise measurement of the $\mu^--\,e^-$ differential cross section, assuming that QED model-independent radiative corrections are under control.

\begin{figure}[t]
\centering
\includegraphics[width=0.3\textwidth]{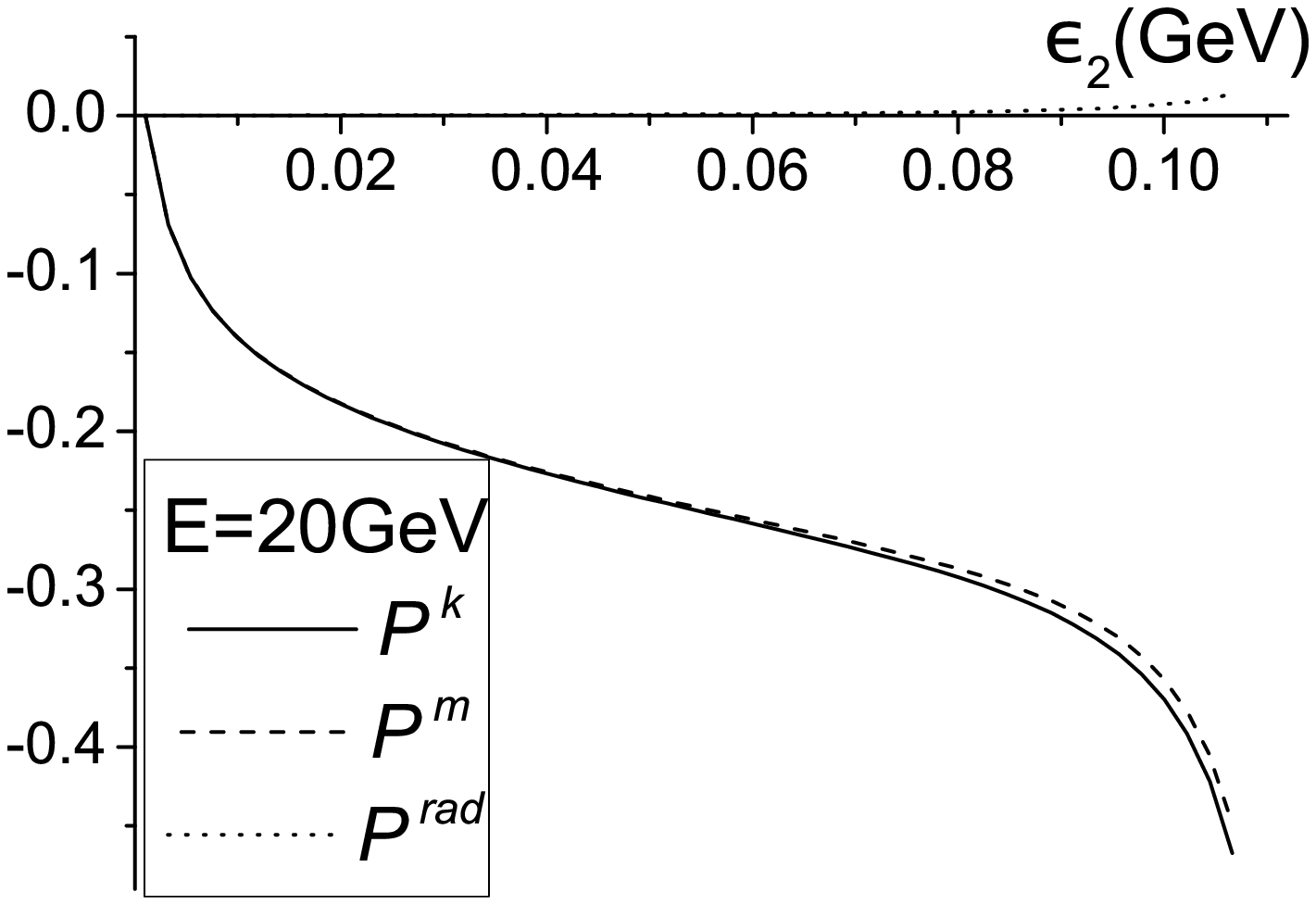}
\includegraphics[width=0.3\textwidth]{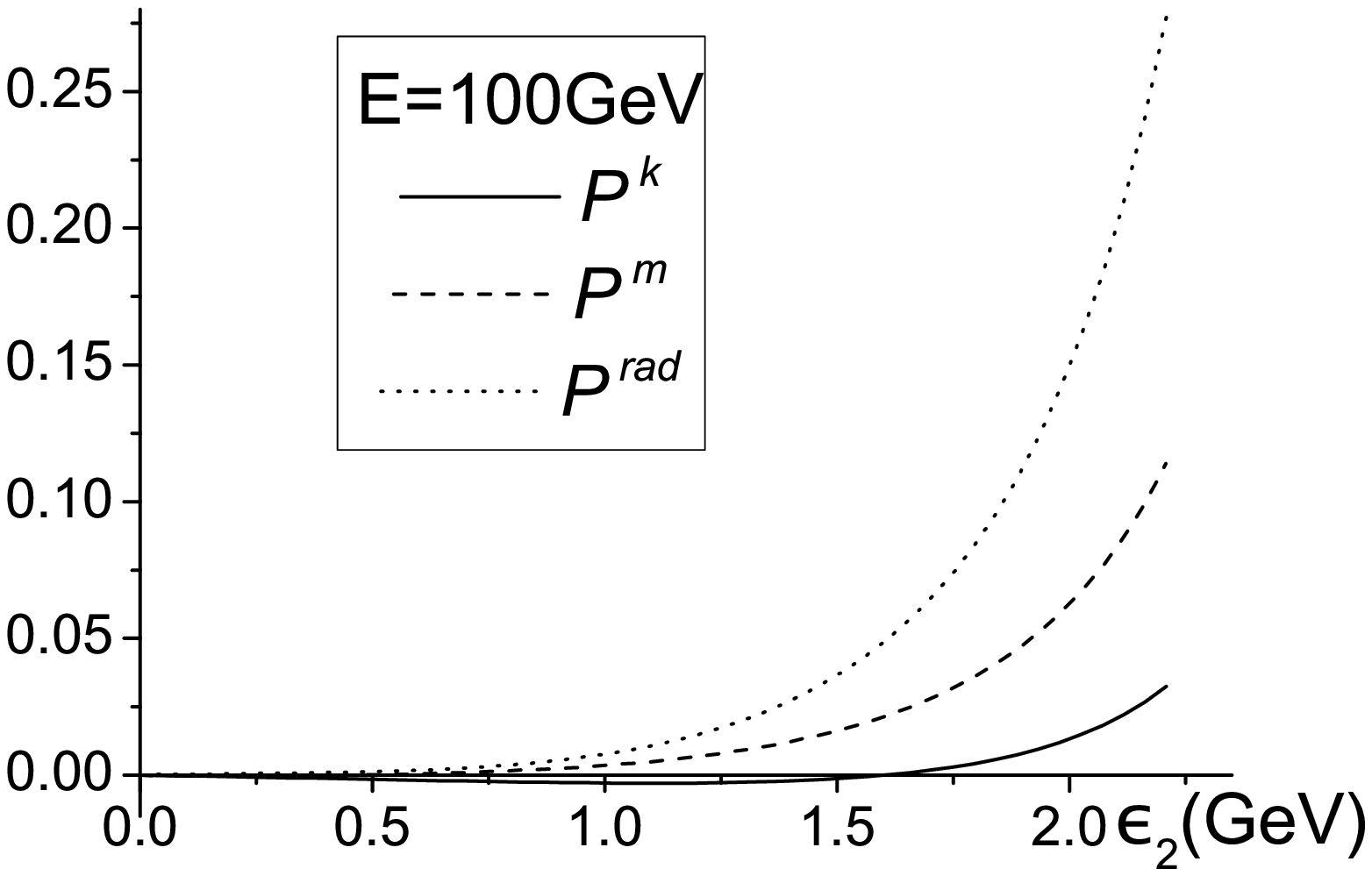}
\includegraphics[width=0.3\textwidth]{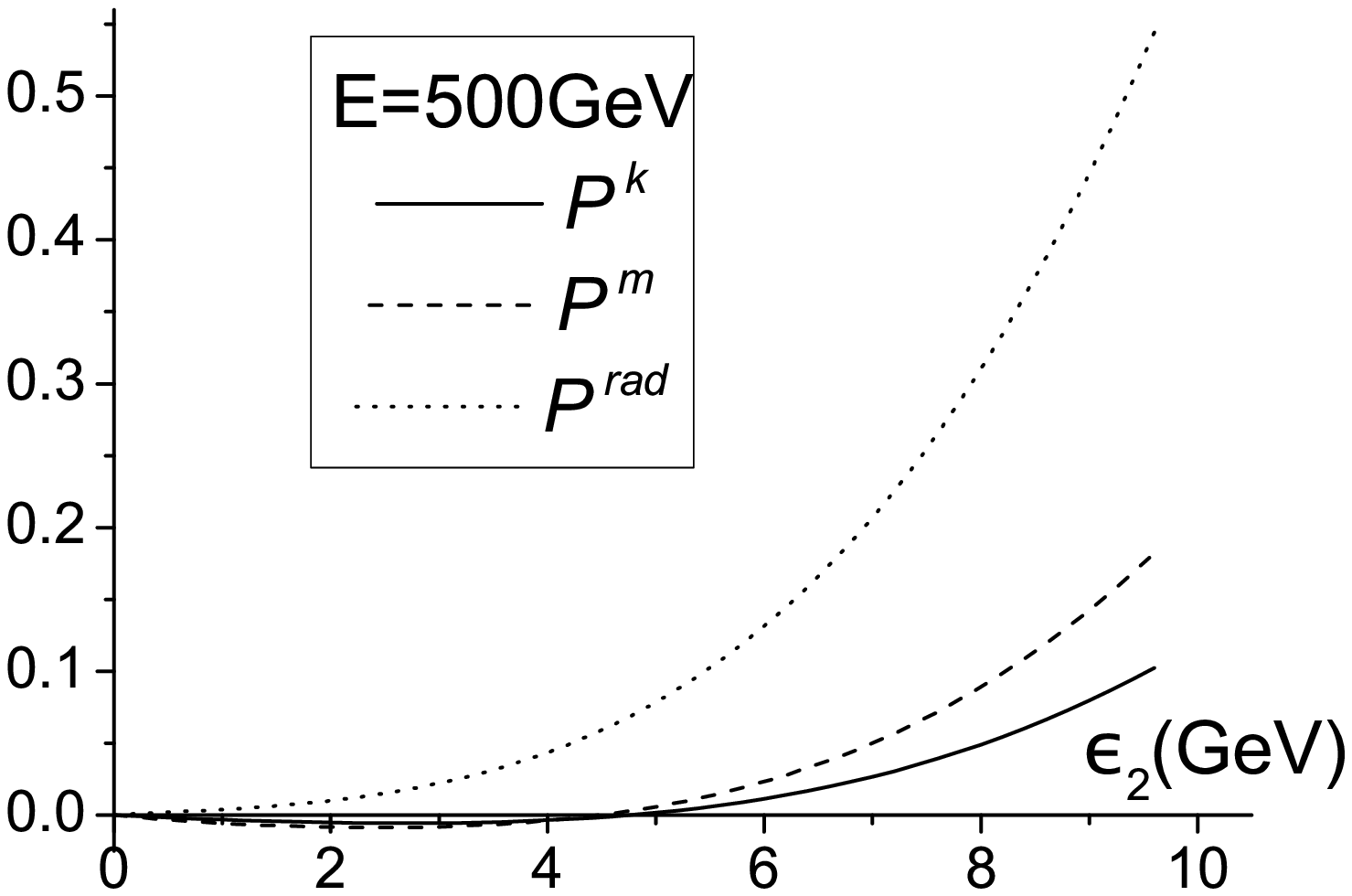}
\caption{Sensitivity of the total model-independent radiative correction (in percent) to the choice of the form factor parametrization, Eq. (\ref{eq:60}).}
 \label{Fig:fig10}
\end{figure}

In Fig. \ref{Fig:fig10} we illustrate the sensitivity of the total radiative correction to the parametrization of the form factors in terms of the ratios
\be\label{eq:60}
P^i=\frac{1+\delta_{\rm tot}^i}{1+\delta_{\rm tot}} -1\,, \ \ i=k,\,m,\,rad\,,
\ee
where $\delta_{\rm tot}$ is the total correction for standard $t_{20}$ fit.
We see that, in the considered conditions,  the deviation of these quantities from unity is very small. We conclude that the influence of the parameterizations of the form factors on the radiative corrections is much smaller than on the Born cross section.

% Moreover, the $r$ and $m$ parameterizations decrease the Born cross section relative to the $d$ one,
%as it follows from Fig.\,10, whereas for the radiative correction we have just opposite effect.

\section{Conclusion}

In this paper we investigated the recoil-electron energy
distribution in elastic deuteron-electron scattering in a coincidence
experimental setup, taking into account the model-independent QED radiative
corrections. The detection of the recoil electron in
this process, with energies from a few MeV up to 10 GeV,
allows to collect small-$Q^2$ data, at 10$^{-5}$\,GeV$^2$
$\leq Q^2\leq$10$^{-2}$\,GeV$^2.$ Such data, being combined
with the existing and  future experiments with electron beams, will give precise information on the
small-$Q^2$ behavior of the deuteron electromagnetic form factors. This 
allows to reach a meaningful extrapolation to the static point and to
extract the deuteron charge radius.
% As noted in the recent review
%\cite{Hill:2017puzz}, it is interesting to extract the proton charge
%radius entirely from low-$Q^2$ data. High precision measurements, in
%the inverse kinematics, allow to accumulate a lot of such data.

To cover the above mentioned interval of  $Q^2$-values, it is
desirable to use the deuteron beams with quite large energies, of the
order of a few hundreds GeV. The sensitivity of
the differential cross section to the form factors parameterizations,
labeled as {\it k, m,} and {\it 20},
is small (does not exceed 2$\%$ ), but the {\it rad}-parametrization  gives a value of the 
cross section about 10$\%$ smaller as compared with  the {\it 20}-parametrization
at $Q^2\approx$\,10$^{-2}$\,GeV$^2$ (see Fig. \ref{Fig:R}).

We took into account the first order QED corrections due to the vacuum
polarization and the radiation of the real and virtual photons by
the initial and final electrons, paying special attention to the
calculation of the hard photon emission contribution when the final
deuteron and electron energies are determined. This hard radiation
takes place due to the uncertainty in the measurement of the deuteron
(electron) energy, $\Delta E_2\, \ (\Delta \varepsilon_2)$. In our
calculations we follow Ref.\,
%\cite{Kahane:1964zz}
\cite{K64}
 for the choice of the
coordinate system and the angular integration method.
Analytical  expressions for the functions
$C_1(\omega)$ and $C_2(\omega),$ defined by Eq. (\ref{eq:46}) can be reconstructed using
the corresponding results for the proton$-$electron scattering which were previously  published \cite{GKM17}. The
cancellation of the auxiliary infrared parameter $\bar{\omega}$ in
the sum of the soft and hard corrections was performed analytically
and the remaining $\omega-$integration in (\ref{eq:45}) was done
numerically.

We assumed that uncertainties in the final particle energies are proportional to their energies and
we showed that the effect due to the nonzero quantity $\Delta \varepsilon_2$ is negligible.
%The increase of parameter $\Delta E$ leads to ??? (Misha comments).

As usual, there is a strong cancellation between the positive hard
correction and negative virtual and soft ones, as it is seen in
Fig. \ref{Fig:fig9}. Despite the fact that the absolute values of these
corrections reach separately 20\,$-$30$\%$, their sum $|\delta_{\rm
tot}|$ does not exceed 6$\%$ at E=20\,GeV and 100\,GeV and 3.5$\%$ at E=500 GeV for
the value $\Delta E_2 =0.02\,(E-\epsilon_2)$ and the $t_{20}$-parametrization used in these calculations.

The total
correction shows a weak dependence on the form factors
parametrization in the considered region  (see. Fig. \ref{Fig:fig10}). At the
lower values of $Q^2$, which correspond to the lower values of the
recoil electron energy $\varepsilon_2,$ the total correction
$\delta_{\rm tot}$ is positive and changes sign when $Q^2$
increases. Such behaviour of $\delta_{\rm tot}$ is similar to the
one found in Ref.\, \cite{K64} and confirmed in Ref. 
\cite{Bardin:1969}
for the case of pion electron scattering.

In our earlier work \cite{GKMT} about proton-electron scattering we have estimated
also the model-dependent part of radiative correction and found that it cannot affect
the experimental cross sections measured within 0.2\,$\%$ accuracy.
We expect that in the case of the deuteron-electron scattering it is even less essential
because the deuteron mass is two times larger.

Thus, we conclude that the model-independent part of the radiative
corrections are under control and, if necessary, it can be
calculated with a more high accuracy. We believe also that the uncertainty
due to its model-dependent part in the considered region is negligible.

%\newpage

%%%%%%%%%%%%%%%%%%%%%%%%%%%%%%%%%%%%%%%%%%%%%%%%%%%%%%%%%%%%%%%%%%%%%%%%%%%%%%%%%%%%

\section*{Acknowledgments}
%%%%%%%%%%%%%%%%%%%%%%%%%%
This work was partially supported by the Ministry of Education and
Science of Ukraine (projects no. 0115U000474 and no. 0117U004866).
The research is conducted in the scope of the IDEATE International Associated Laboratory (LIA).
%%%%%%%%%%%%%%%%%%%%%%%%%%%%%%%%%%%%%%%%%%%%

\end{document}